\documentclass[12pt]{iopart}
  \expandafter\let\csname equation*\endcsname\relax
  \expandafter\let\csname endequation*\endcsname\relax
\usepackage{amsmath}
\usepackage{bbm}
\usepackage{bm}
\usepackage{latexsym}
\usepackage{graphicx}
\usepackage{epstopdf}
\usepackage{color} 
\usepackage{subfigure}
\usepackage{amsfonts}
\newcommand{\beq}{\begin{equation}}
\newcommand{\eeq}{\end{equation}}

\begin{document}

\title{A simple non-equilibrium, statistical-physics toy model of thin-film growth}

\author{Jeremi Kazimierz Ochab$^1$, Hannes Nagel$^2$, Wolfhard Janke$^2$, Bart{\l}omiej Waclaw$^3$}
\address{
	$^1$Marian Smoluchowski Institute of Physics, Jagiellonian University, ul. S. \L{}ojasiewicza 11, 30-348 Krak\'ow, Poland\\
  $^2$Institut f\"ur Theoretische Physik, Universit\"at Leipzig, Postfach 100\. 920, D-04009 Leipzig, Germany\\
  $^3$School of Physics and Astronomy, University of Edinburgh, James Clerk Maxwell Building, Peter Guthrie Tait Road, Edinburgh EH9 3FD, United Kingdom}
\eads{\mailto{jeremi.ochab@uj.edu.pl}}

\begin{abstract}
We present a simple non-equilibrium model of mass condensation with Lennard-Jones interactions between particles and the substrate. We show that when some number of particles is deposited onto the surface and the system is left to equilibrate, particles condense into an island if the density of particles becomes higher than some critical density. We illustrate this with numerically obtained phase diagrams for three-dimensional systems. We also solve a two-dimensional counterpart of this model analytically and show that not only the phase diagram but also the shape of the cross-sections of three-dimensional condensates qualitatively matches the two-dimensional predictions. Lastly, we show that when particles are being deposited with a constant rate, the system has two phases: a single condensate for low deposition rates, and multiple condensates for fast deposition. The behaviour of our model is thus similar to that of thin film growth processes, and in particular to Stranski-Krastanov growth.
\end{abstract}

\pacs{89.75.Fb, 05.40.-a, 64.60.Ak}
Date \today

\maketitle

\section{Introduction}

Non-equilibrium statistical mechanics has witnessed a rapid progress in recent years, and has been applied to a variety of problems in physics, chemistry, biology, economy, and social sciences. However, in contrast to equilibrium systems, which can be conveniently studied by using the concept of the statistical ensemble, a unified theoretical framework applicable to all non-equilibrium systems does not exist, and whether such a framework will eventually emerge remains to be seen. 

Despite that, significant progress has been made in the last two decades for a class of models called driven diffusive systems~\cite{DDS} which -- even though being far from equilibrium -- can be studied within the same statistical ensemble framework as equilibrium models. These models share a common feature: the steady-state probability of a microstate can be expressed analytically as a function of transition rates which define the dynamics of the model. Examples of such systems are the zero-range process (ZRP)~\cite{Evans_2000,Evans_JPA2005,groskinsky_condensation_2003}, closely related to its equilibrium counterpart: balls-in-boxes model (B-in-B)~\cite{Bialas}, the asymmetric simple exclusion process (ASEP)~\cite{ASEP} and its totally asymmetric version (TASEP)~\cite{derrida_exact_1993}, asymmetric inclusion process (ASIP)~\cite{ASIP, grosskinsky_condensation_2011, cao_dynamics_2014} and many variations on these two models~\cite{Evans_JPA2004, hirschberg_motion_2012, godreche_condensation_2012, godreche_urn_2007, daga_phase_2015}. In all these models, particles jump between sites of a one- or higher-dimensional lattice and the dynamics is defined by specifying the hopping rates of the particles. The hopping rates are usually chosen so that there is a non-zero, macroscopic current of particles through the system driving it far from equilibrium, although the system often exhibits a non-equilibrium steady state independent of the initial condition.

In this paper, we study an extension of the zero-range process to nearest-neighbour interactions, similar to that of Refs.~\cite{Evans_PRL2006,BW_PRL}. In this model, particles interact when they are at the same site or at neighbouring sites. Although the model can be driven far from equilibrium, it is closely related to the equilibrium solid-on-solid (SOS) model~\cite{SOS1,SOS2,SOS3}. A remarkable feature of this stochastic process is that the steady state factorises over \emph{pairs} of neighbouring sites, also in dimensions higher than one, and thus we call it the pair-factorised steady state process (PFSS). This property facilitates analytical calculations in the one-dimensional version of the model, and in certain cases also in more than one dimension~\cite{BW_JPA2009}.

\begin{figure}
	\centering
		\includegraphics[width=0.32\textwidth]{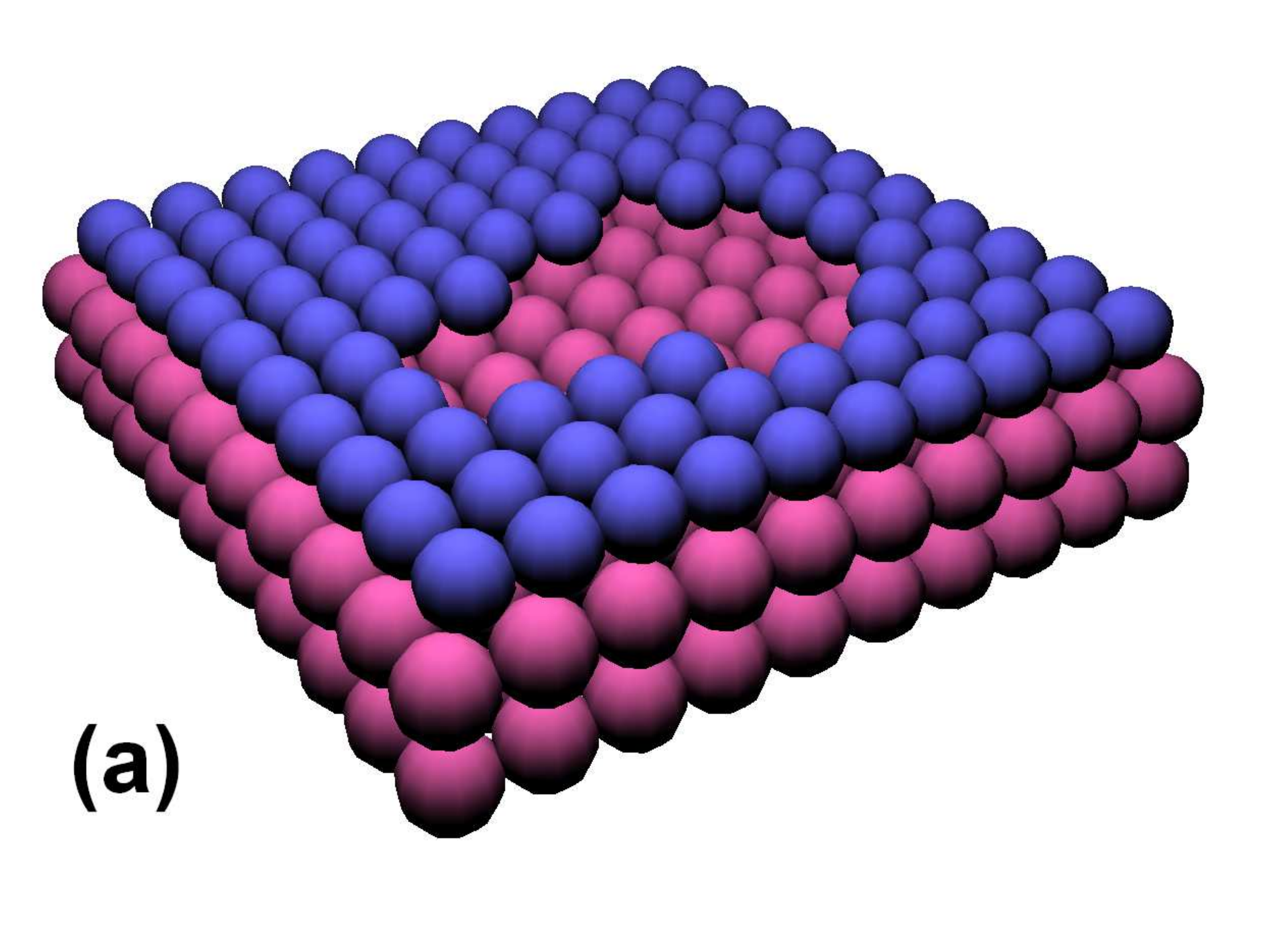}
		\includegraphics[width=0.32\textwidth]{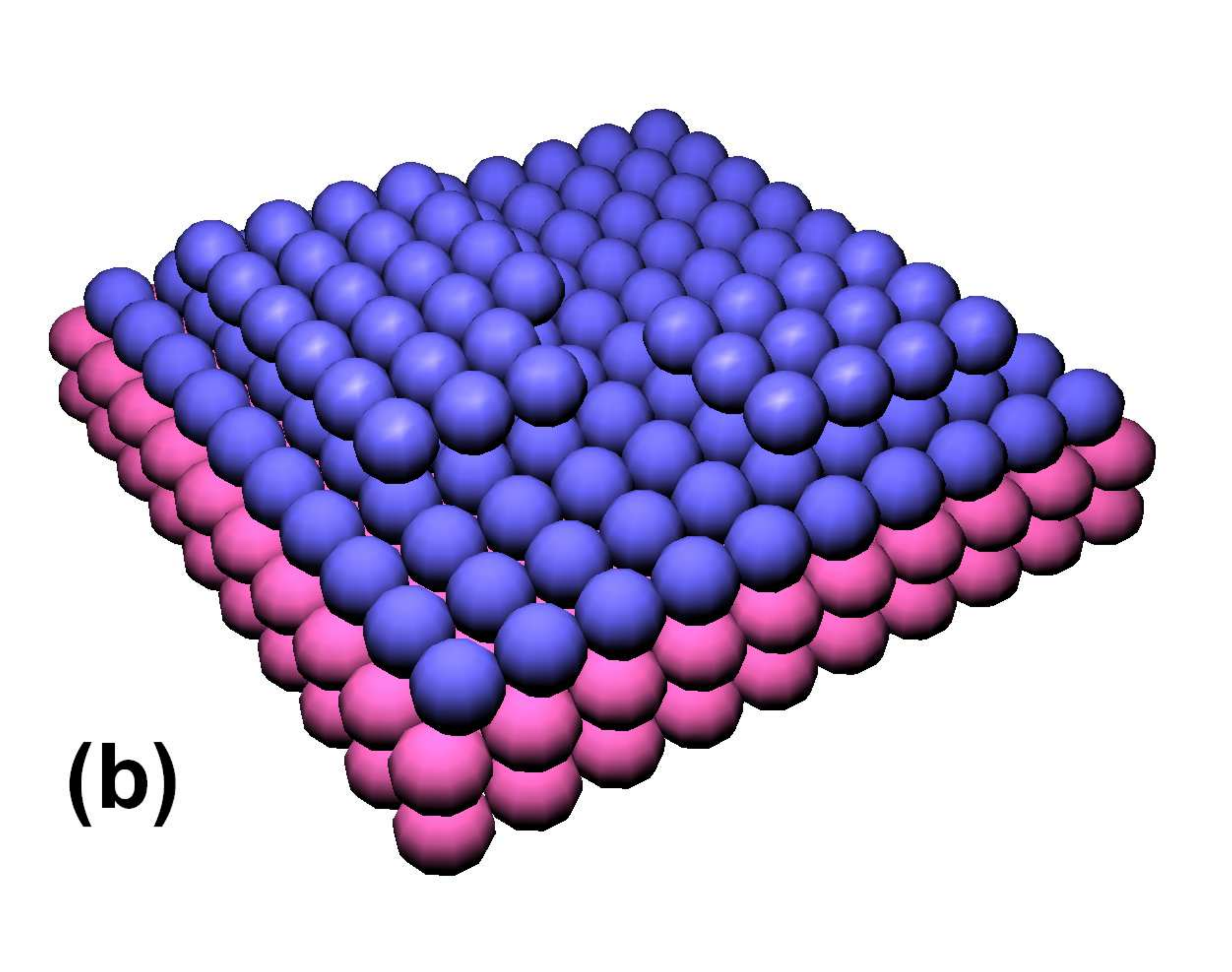}
		\includegraphics[width=0.32\textwidth]{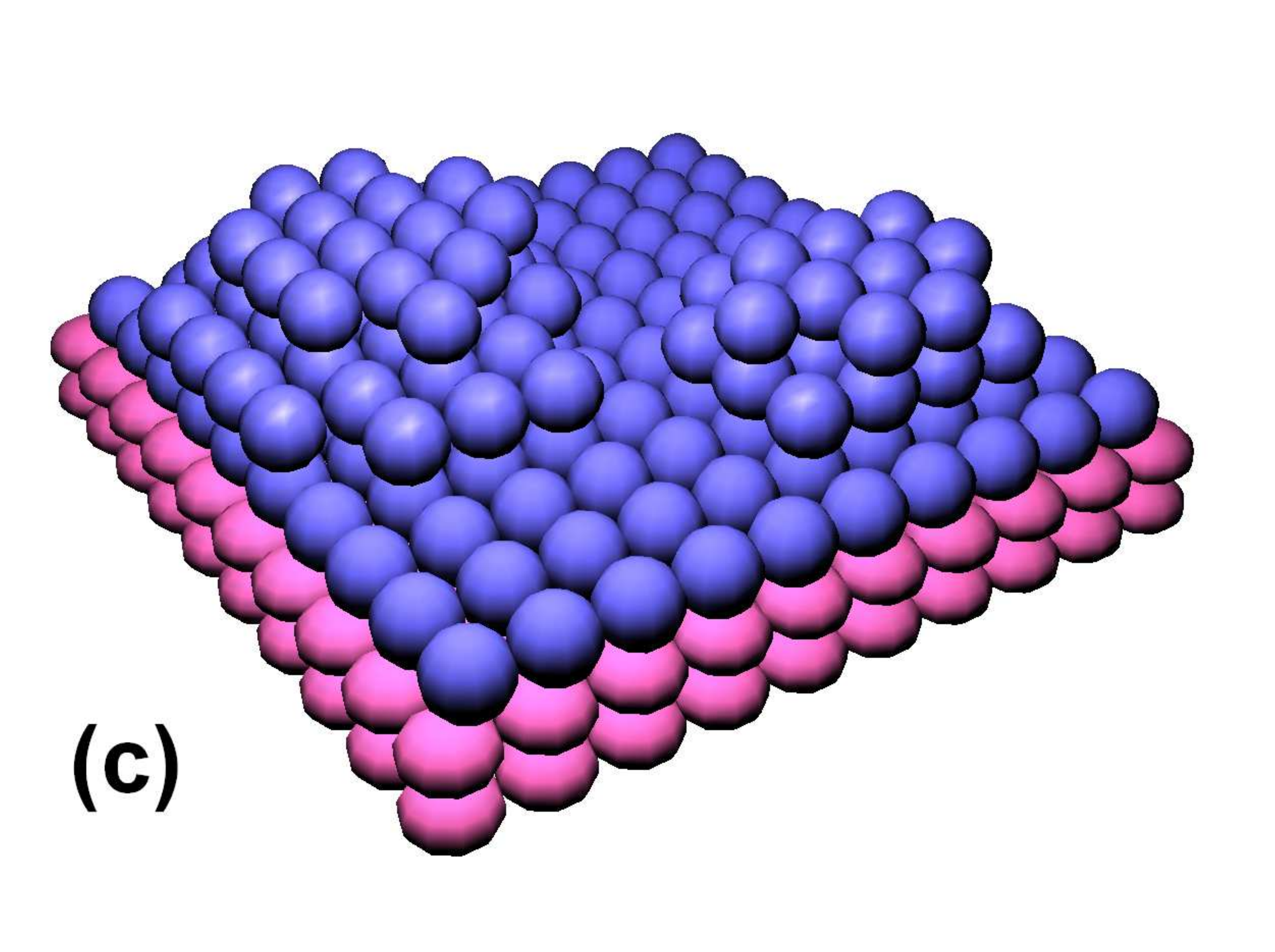}
	\caption{\label{fig:SKgrowth} Schematic stages of Stranski-Krastanov growth. Adatoms (blue spheres) are deposited on the substrate (purple spheres) until a desired density is reached. (a) Low density -- an incomplete monolayer is formed. (b) As the density of adatoms increases, the adatoms form a complete monolayer and a partially filled second layer. (c) Upon further density increase, islands of variable height begin to form on the monolayer. Here the critical density of adatoms above which islands occur equals one adatom per one substrate site. The critical thickness depends on the mismatch between the substrate and adsorbate lattices, which is, however, not modelled explicitly in this paper.
}
\end{figure}

Contrary to previous works which focused on generic properties of this model such as the existence of condensation~\cite{Evans_PRL2006,BW_JStat}, the shape of the condensate~\cite{BW_PRL}, or generalisation to more complicated graphs~\cite{BW_JPA2009}, we revisit here the original foundation of this model coming from solid-state physics, and choose a hopping rate which leads to the emergence of clusters of particles similar to the extended atomic islands known from non-equilibrium nanostructure formation~\cite{Meakin_book} and epitaxial thin film growth~\cite{Krug2}. In these processes, a film of atoms is deposited on a substrate that serves as a template. One of three generic modes of epitaxial thin film growth~\cite{venables_introduction_2000} -- Stranski-Krastanov growth -- has attracted considerable attention as it can be used, for example, to produce quantum dots~\cite{Qdots,Qdots1}. In Stranski-Krastanov growth, deposited atoms form initially a flat, 2d layer. As the density of atoms on the substrate increases beyond a certain critical thickness, atomic islands start to nucleate as shown schematically in Fig.~\ref{fig:SKgrowth}. 

In this paper, we propose a simple, analytically tractable non-equilibrium toy model that mimics the 2d-to-3d transition observed in Stranski-Krastanov growth. In our model, we do not aim at reproducing all details of thin film growth (e.g., there is no mismatch between the substrate and adsorbate lattices) but we rather explore generic mechanisms that lead to island formation in non-equilibrium systems that mimic those of Stranski-Krastanov mode of growth. In particular, we show that by changing the strength of interactions between particles one can obtain different island shapes, similarly to what is seen in experiments. We also show that the shape is quite robust and does not change much when the system is pushed far from equilibrium either by imposing a macroscopic current of particles in one direction (as in electromigration on surfaces~\cite{kandel_microscopic_1996}), or by adding new particles to the system at a constant rate (the latter process imitating molecular beam epitaxy~\cite{Krug2}).

\section{The model}
\label{sec:pfss}

The model that we study in this work comprises a two-dimensional, regular lattice with $N=L\times L$ sites and periodic boundary conditions in both directions. Let $\{m_i\}$ be the number of particles at sites $i=1,\dots,N$. The particles can be viewed as ``adatoms'' attached to the surface of the substrate where the number $m_i$ corresponds to the height (in the third dimension) of a stack of atoms at the site $i$. We first consider the case when $M$ particles have been deposited on the substrate and no further particles are being added, thus the number of particles is constant and equal to $M$. We shall later relax this assumption. 

To model the dynamics of particles due to thermal excitations and external driving (e.g. electromigration), we assume the following rate at which a particle jumps out of site $i$:
\beq
	\label{eq:Um}
	u_{i}=\prod_{\langle i,j\rangle} \frac{g(m_{i}-1,m_{j})}{g(m_{i},m_{j})},
\eeq
where $\langle i,j\rangle$ denotes all four nearest neighbours of site $i$ and $g(m,n)$ is a symmetric non-negative function to be specified later. The jump rate depends on the number of particles at $i$ and at its nearest neighbours, and by a suitable choice of $g(m,n)$ we can replicate interactions between particles at neighbouring sites.
The particle then hops to one of the neighbours with probabilities $\{r_k\}$ for $k=1$~(right), $2$~(left), $3$~(top), and $4$~(bottom). The above choice of $u_i$ is dictated by the requirement that the steady-state microstate probability assumes the following factorized form~\cite{BW_JStat},
\beq
\label{eq:Pm}
P(m_1,\ldots,m_N)= \frac{1}{Z}\prod_{\langle i,j\rangle}{g(m_i,m_{j})\delta\left[\sum_{i=1}^{N}{m_i}-M\right]},
\eeq
where the probability factorises over pairs of neighbouring sites, the Kronecker delta $\delta[k]$ (equal to 1 if $k=0$) ensures that the total number of particles is conserved, and $Z$ is a normalisation constant giving (\ref{eq:Pm}) a valid probabilistic interpretation.
The factorisation allows us to analyse the statics of the model using standard tools of statistical mechanics.
Identifying $P(m_1,\ldots,m_N)$ with the Boltzmann distribution $(1/Z)\exp(-\beta E)$ with the inverse temperature $\beta=1$, we obtain the energy of the microstate
\beq
	E(m_1,\ldots,m_N) = -\sum_{\langle i,j\rangle} \ln g(m_i,m_{j}).
\eeq
Even though the system is in general out of equilibrium, the steady state is independent of the jump probabilities $\{r_k\}$, and many steady-state quantities can be calculated as if the system was in equilibrium, with the microstate probability given by Eq.~(\ref{eq:Pm}). The choice of the probabilities $\{r_k\}$ determines how far the system is from equilibrium; for example, for $\{r_k\}=\{1/3,0,1/3,1/3\}$ particles jump asymmetrically from left to right, which generates a macroscopic current of particles in this direction, whereas for $\{r_k\}=\{1/4,1/4,1/4,1/4\}$ the jumps are fully symmetric, the net current of particles is zero, and the system is at equilibrium.

We also consider a (1+1)d counterpart of this model, in which particles jump to the right or left on a one-dimensional substrate, and the simplified form of the hopping rate Eq.~(\ref{eq:Um}) is
\beq
\label{eq:Um1d}
u(m_i|m_{i+1},m_{i-1})=\frac{g(m_{i}-1,m_{i-1})}{g(m_{i},m_{i-1})}\frac{g(m_{i}-1,m_{i+1})}{g(m_{i},m_{i+1})}.
\eeq 
The corresponding microstate probability then reads
\beq
\label{eq:Pm1d}
P(m_1,\ldots,m_N)= \frac{1}{Z}\prod_{i=1}^{N}{g(m_i,m_{i+1})\delta\left[\sum_{i=1}^{N}{m_i}-M\right]}.
\eeq 
Similarly to the (2+1)d model, the probability $r_1=r_2=1/2$ corresponds to the system in thermal equilibrium, whereas for $r_1=1,r_2=0$ the particles can jump only to the right as in Ref.~\cite{Evans_PRL2006}.

The model described above has been studied for a number of choices of $g(m,n)$ in one dimension~\cite{Evans_PRL2006,BW_PRL}, and less extensively in two dimensions~\cite{BW_JPA2009} and, given that $g(m,n)$ fulfils certain criteria, the model is known to have a phase transition between a liquid and a condensed state as the density of particles crosses a threshold density. Depending on the choice of $g(m,n)$, the condensate can be either localised at a single site as in the ZRP, or can be spatially extended over many sites~\cite{BW_PRL,ehrenpreis_numerical_2014}.

\begin{figure}
	\centering
        \includegraphics[width=0.4\textwidth]{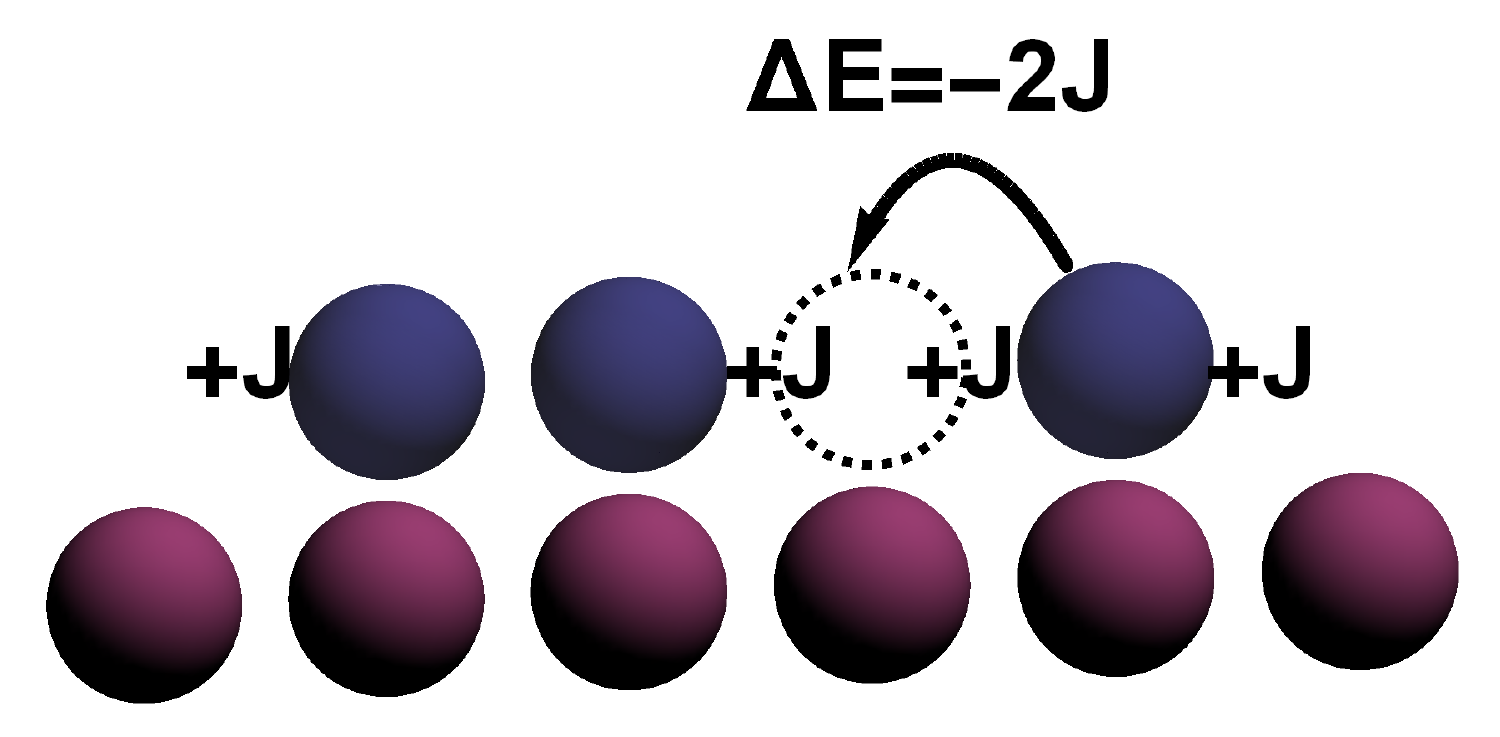}
        \hspace{2cm}
        \includegraphics[width=0.4\textwidth]{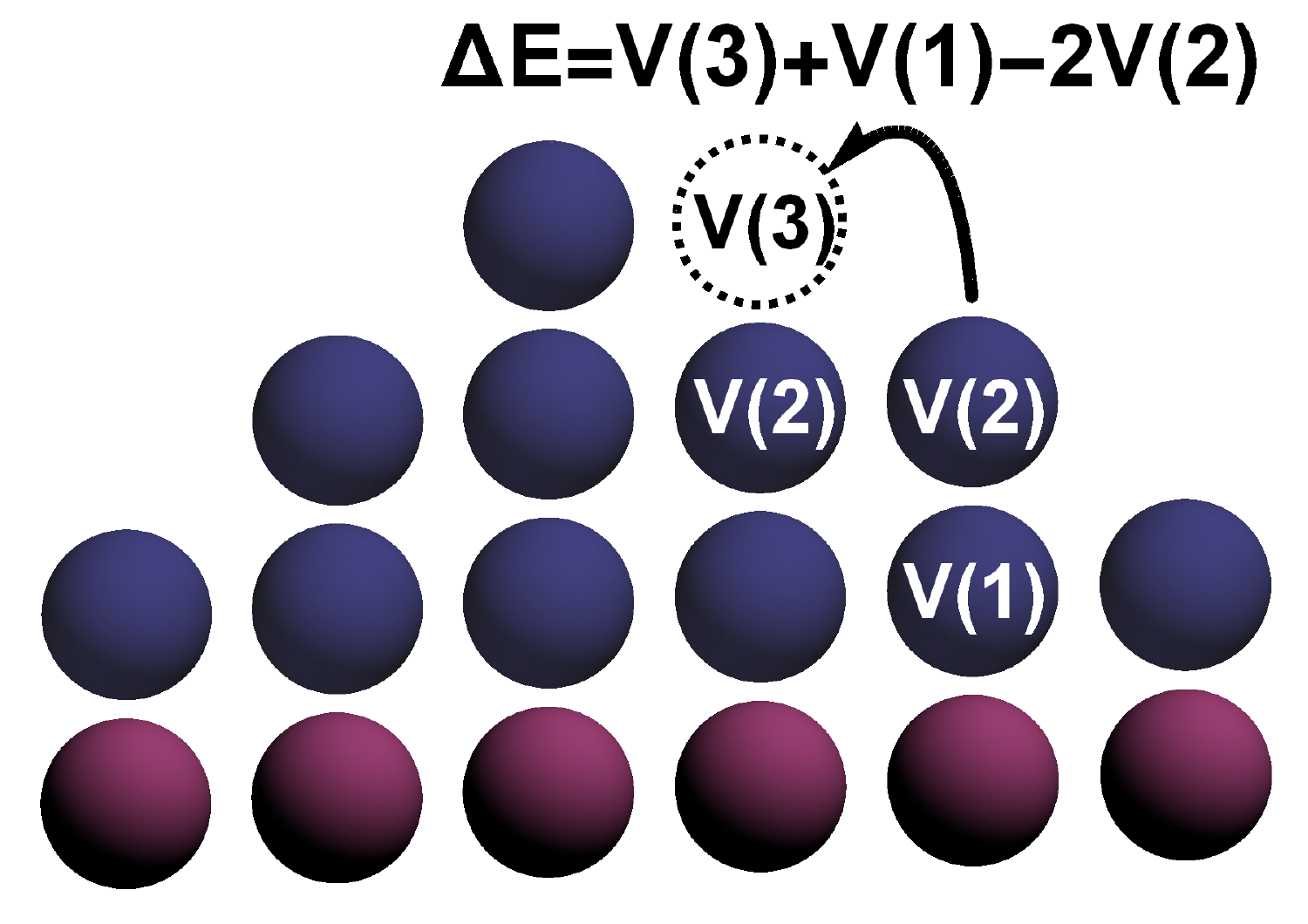}
	\caption{\label{fig:hops} When a particle jumps to a neighbouring site with rate (\ref{eq:Um1d}), the energy of the system changes. Left: a particle merging with a two-particle island reduces the number of height differences (``broken bonds'') $\sum_i |m_i-m_{i+1}|$ by two;
	right: a particle hopping onto a neighbouring stack of adatoms changes the value of the on-site potential $V(m)$.
	For $\sigma=1, J>0, U>0$ both depicted energy changes are negative, $\Delta E<0$, hence these transitions are more likely than the moves in the opposite direction.
}
\end{figure}

In this paper, we aim to reproduce qualitatively the phenomenology of surface growth. We therefore assume the energy of a configuration $(m_1,\ldots,m_N)$ to be
\beq
	E = J \sum_{\langle i,j\rangle} |m_i-m_j| + U \sum_i \left[\left(\frac{\sigma}{m_i+1}\right)^9-\left(\frac{\sigma}{m_i+1}\right)^3\right],
\eeq
which is equivalent to the following two-point symmetric weight function $g(m,n)$:
\beq
	g(m,n) = \exp\left[-J|m-n| - \frac{1}{2}(V(m)+V(n))\right], \label{eq:gmndef}
\eeq
where $V(m)$ is the potential 
\beq
\label{eq:LJmodel}
	V(m) = U \left[\left(\frac{\sigma}{m+1}\right)^9-\left(\frac{\sigma}{m+1}\right)^3\right].
\eeq
The term proportional to $J$ in the above expression represents the energy cost of ``broken bonds'' between neighbouring, vertical stacks of adatoms, whereas the term proportional to $U$ accounts for interactions between adatoms and the substrate, see Fig.~\ref{fig:hops}. We assume the latter to be described by the Lennard-Jones (LJ) potential, with a unity added to the denominator to make the expression finite for $m=0$. Equation (\ref{eq:LJmodel}) is relevant to interactions of molecules with a crystalline surface~\cite{LJ93_WS,LJ93_RH}, and the exponents $9$ and $3$ arise from integrating the 12-6 LJ potential over the substrate surface\footnote{The model can be easily extended to other choices of the potential, including other exponents, and will exhibit qualitatively the same behaviour.}.

\begin{figure}
	\centering
		\includegraphics[width=.49\textwidth]{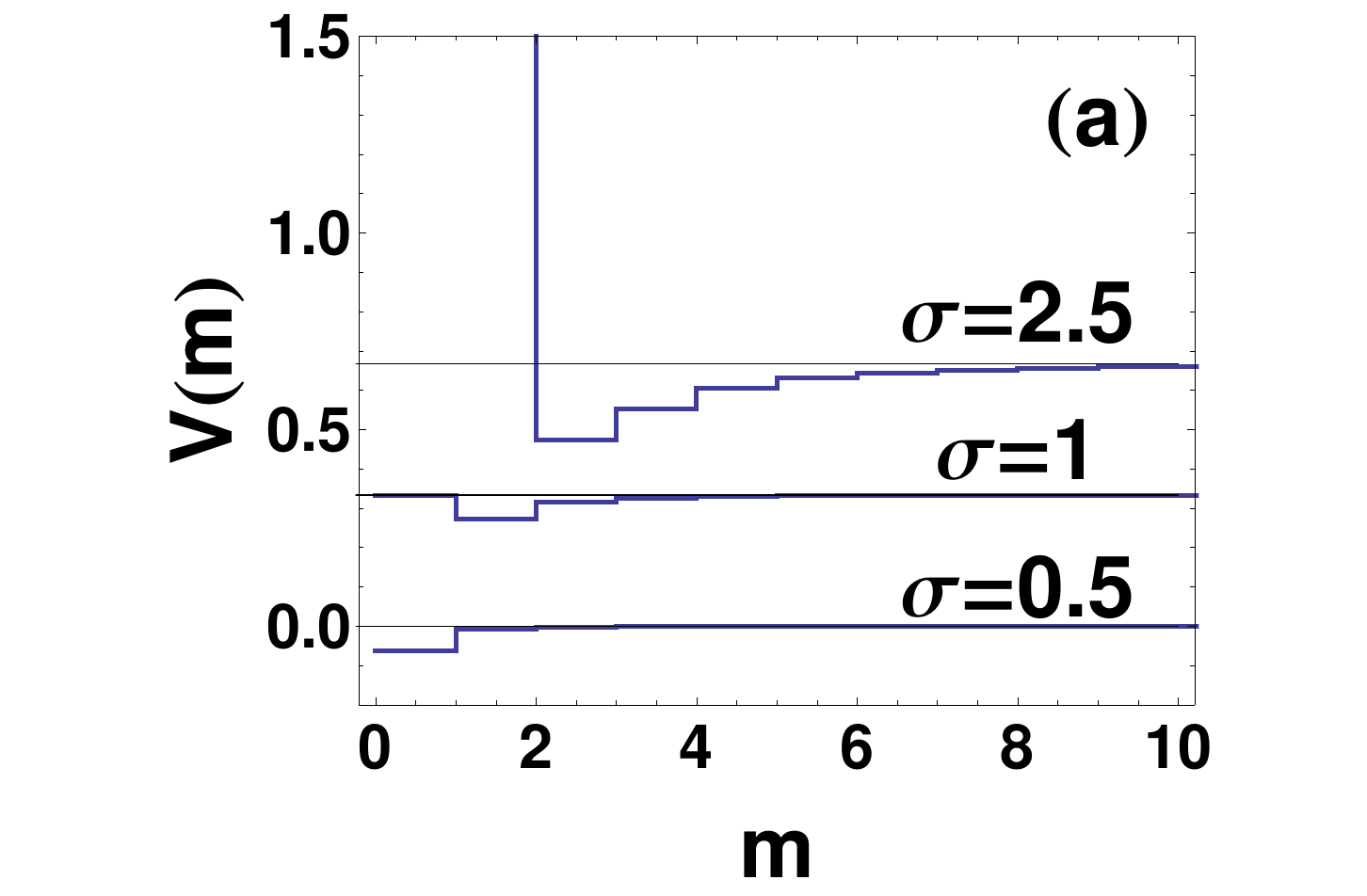}
		\hfill
		\includegraphics[width=.49\textwidth]{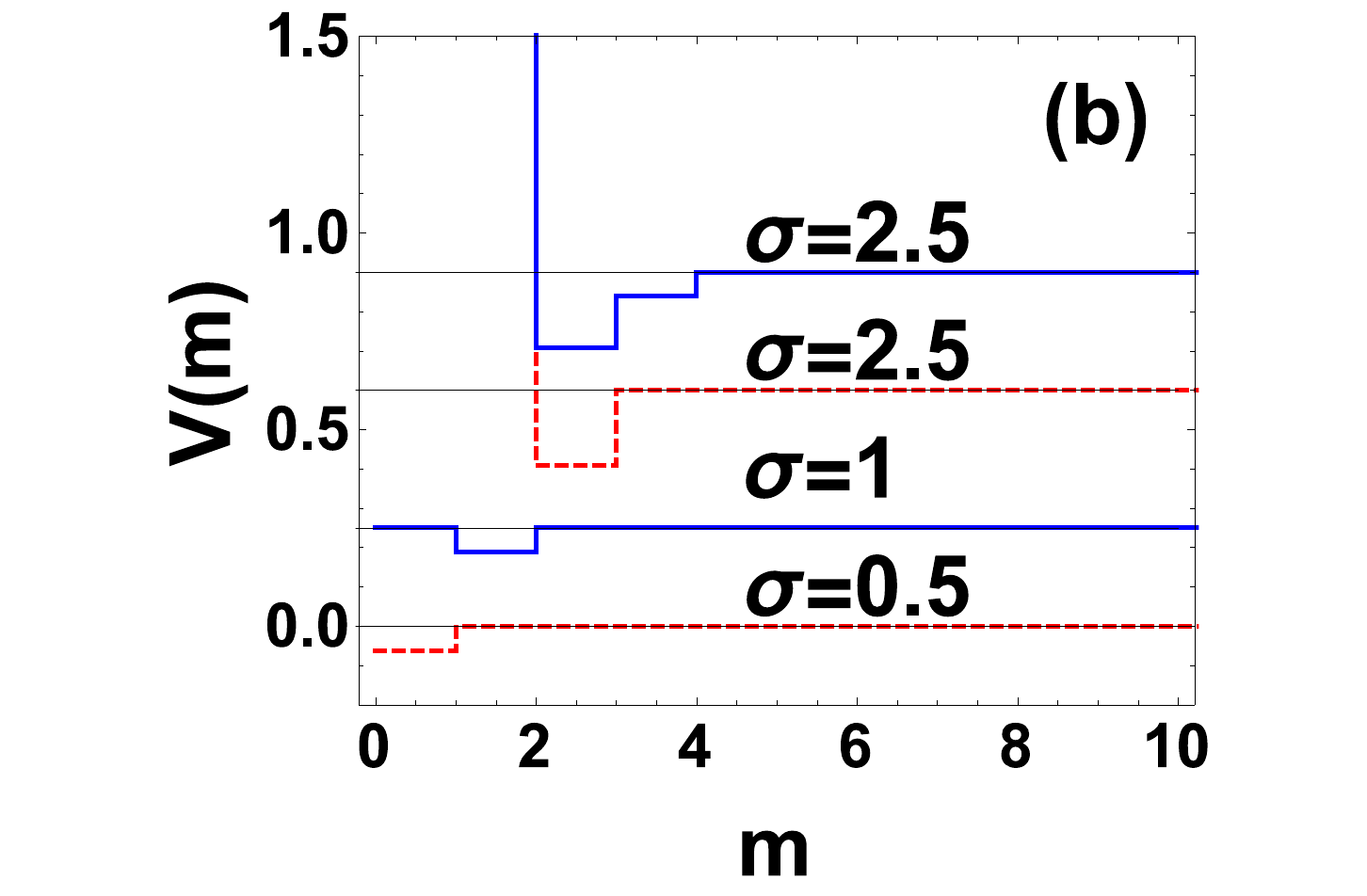}
		\caption{\label{fig:potential} (a) Examples of the on-site potential $V(m)$ of Eq.~\eqref{eq:LJmodel} for $U=1/2$. (b) Approximate potentials for $\sigma=0.5, 1$, and $2.5$ constructed from one (dashed lines) or two (continuous line) delta functions; see Sec.~\ref{sec:modelA}--C. 
The plots have been vertically shifted for clarity; black horizontal lines mark the zero energy level in each case.}
\end{figure}

The model has three parameters that are related to adatom-substrate interactions: $U, J$ and $\sigma$. Large $J$ suppresses, through $|m_{i}-m_{j}|$, rapid variations in the height of neighbouring stacks $\langle i,j\rangle$ of adatoms, and flattens out the surface; large $U$ makes the adatoms bind stronger to the substrate; $\sigma$ has the interpretation of the interaction range between the adatoms and the surface, measured in the units of the lattice constant. Figure~\ref{fig:potential}a shows how the potential $V(m)$ behaves for different values of the parameter $\sigma$, for $U=1/2$. 
Although our model only serves as heuristic means, the values of $\sigma$ we use throughout the paper ($0<\sigma\leq3$) fall in the range typically encountered in Stranski-Krastanov growth, see~\ref{app:atoms}.

\section{Numerical results for the (2+1)d model}
\label{sec:2D}

In this section we discuss steady-state properties of the (2+1)d model with fixed number of particles. Since all quantities discussed here depend only on $g(m,n)$ through the steady-state probability (\ref{eq:Pm}) and not on transition probabilities between the states of the system, we took the liberty of using a Monte Carlo algorithm (for more details, see~\ref{app:comp}) to simulate the (2+1)d model on a computer. This approach, unsuitable for dynamic quantities such as the average current, is much better suited for simulations of large systems due to its significant speed gain over the dynamics described by Eq.~(\ref{eq:Um}).
 
\begin{figure}
	\centering
		\includegraphics[width=0.495\textwidth]{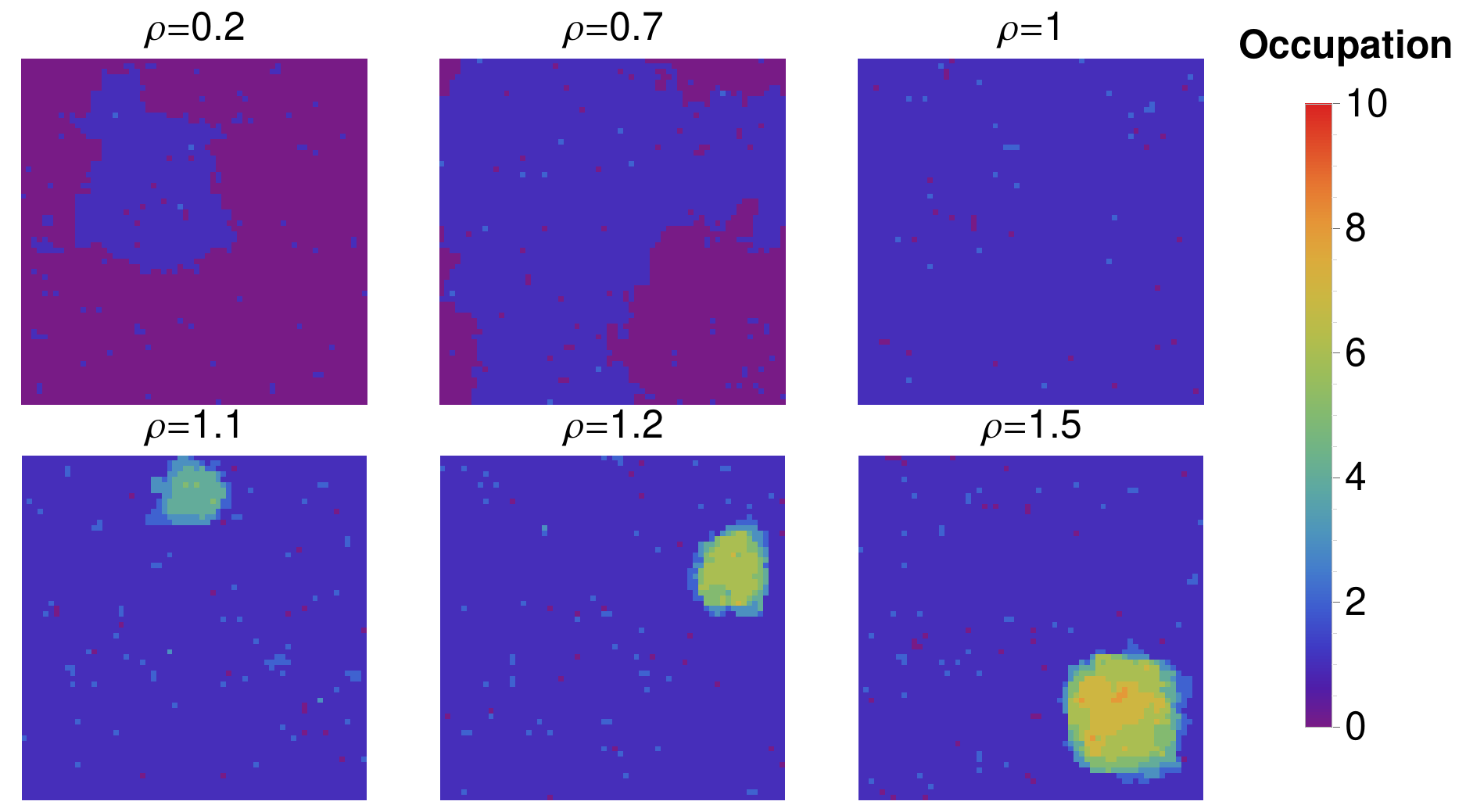}
		\hfill
		\includegraphics[width=0.495\textwidth]{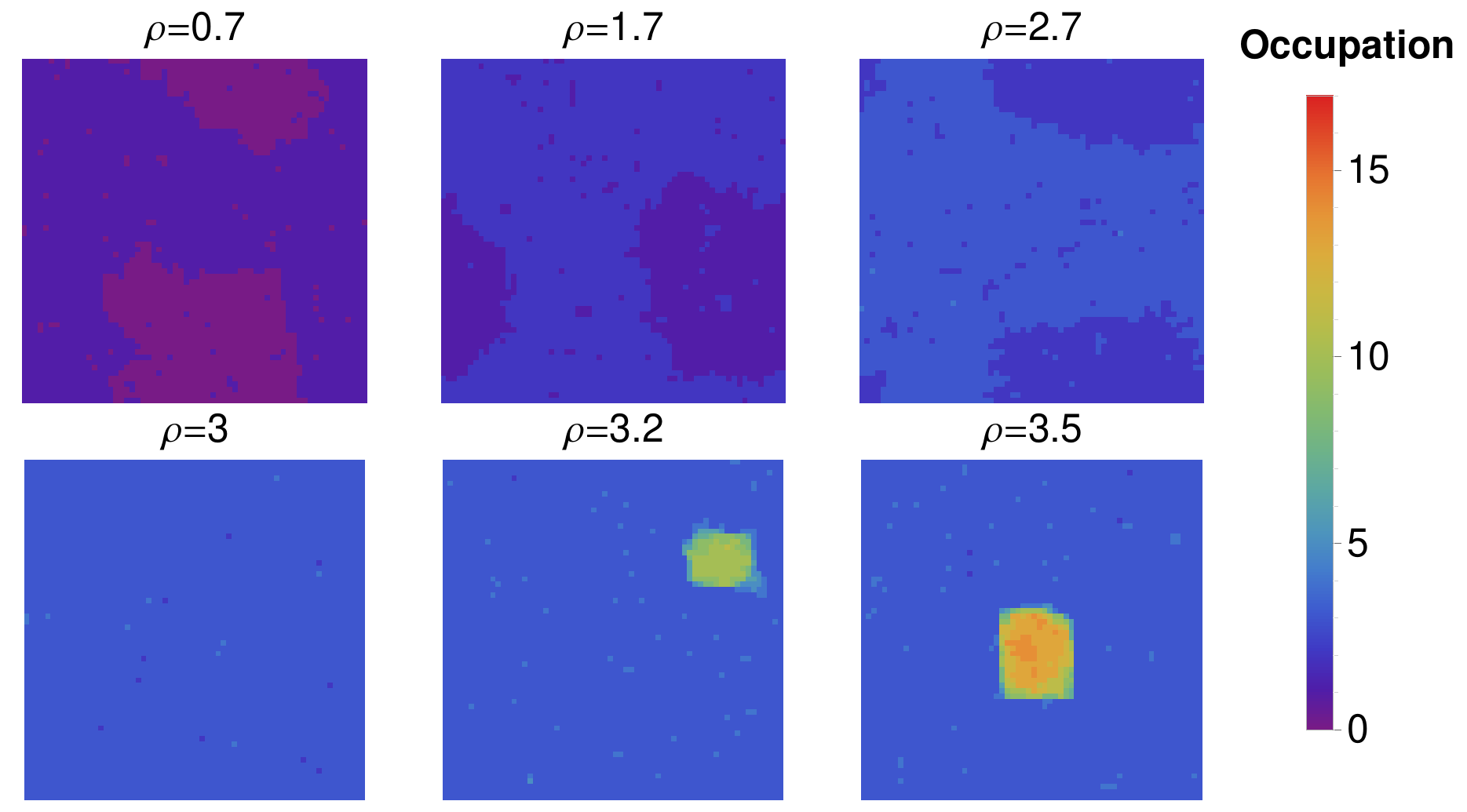}
	\caption{\label{fig:densityS1}(Color online) Snapshots of the steady-state of the system for different densities of particles $\rho=M/L^2$; colours represent different numbers of particles per site (see bar on the right). Each square represents a $64\times 64$ lattice with periodic boundary conditions. The condensate (a green/yellow/red shape) forms for large enough densities.
The parameters are $J=1.1, U=3$, and (left) $\sigma=1$, (right) $\sigma=3$. 
}
\end{figure}

Figure~\ref{fig:densityS1} shows steady-state snapshots of the system for different surface densities $\rho=M/L^2$, fixed $U,J$, and for two different $\sigma=1,3$. For $\sigma=1$, increasing the density $\rho$ produces first a flat, irregular droplet of height $m=1$ and size increasing with $\rho$; then, above a certain critical density $\rho_{\text{c}}$, a hemi-spherical island -- which we shall call the condensate -- begins to form on the surface. The height of the condensate increases with $\rho$, while the height of the surface remains constant and equal to one. The situation looks similar for $\sigma=3$ except that the condensate forms on a layer of three particles thick. The snapshots suggest that the critical density for condensation is approximately equal to $\sigma$, the range of the LJ potential. Indeed, simulations made for $U,J,\sigma$ as in Fig.~\ref{fig:densityS1} and for a range of densities $\rho=1,\dots, 7$ show that $\rho_{\text{c}} = 1.0135\pm0.0013$ for $\sigma=1$ and $\rho_{\text{c}}=3.068\pm0.023$ for $\sigma=3$ (see~\ref{app:rhoc}). The values of $\rho_{\text{c}}$ remain very close to $\left\lfloor \sigma\right\rfloor$ (i.e., the floor of $\sigma$), which indicates that, for $U$ large enough, the critical density is very close to the density of particles necessary to populate the first $\left\lfloor\sigma\right\rfloor$ layers above the substrate. We shall see in Sec.~\ref{sec:1D} that this is also true for the analytically solvable (1+1)d model.

\begin{figure}[htbp]
	\centering
		\includegraphics[width=\textwidth]{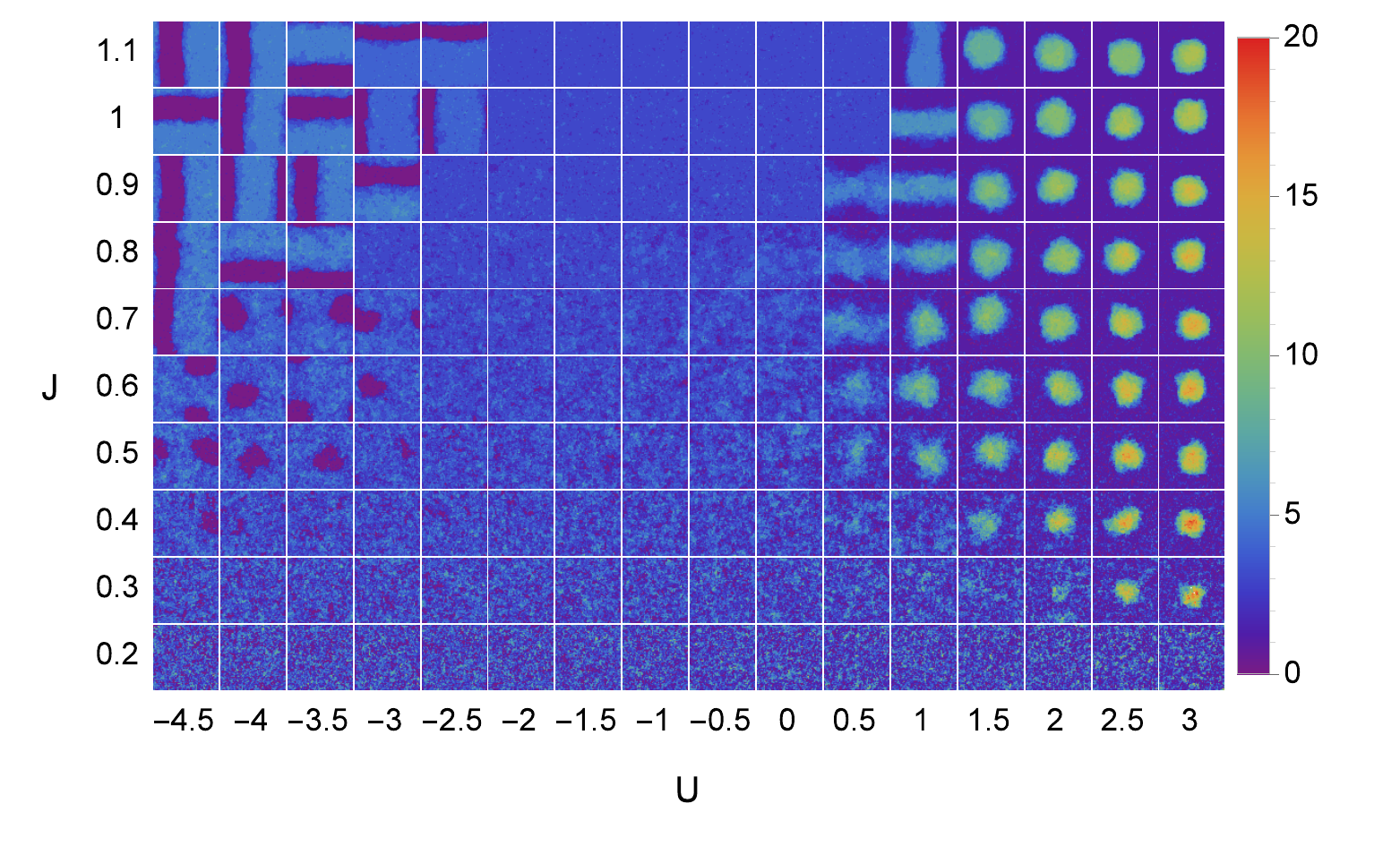}
	\caption{\label{fig:diagramS1} (Color online) Phase diagram for $\sigma=1$; each square represents a snapshot of the (2+1)d system with $64\times64$ sites, simulated for a given pair $U,J$. Colours represent different numbers of particles per site (see the colour bar). The condensates have been shifted so that each appears in the centre of the lattice.}
\end{figure}

\begin{figure}[htbp]
		\includegraphics[width=0.325\textwidth]{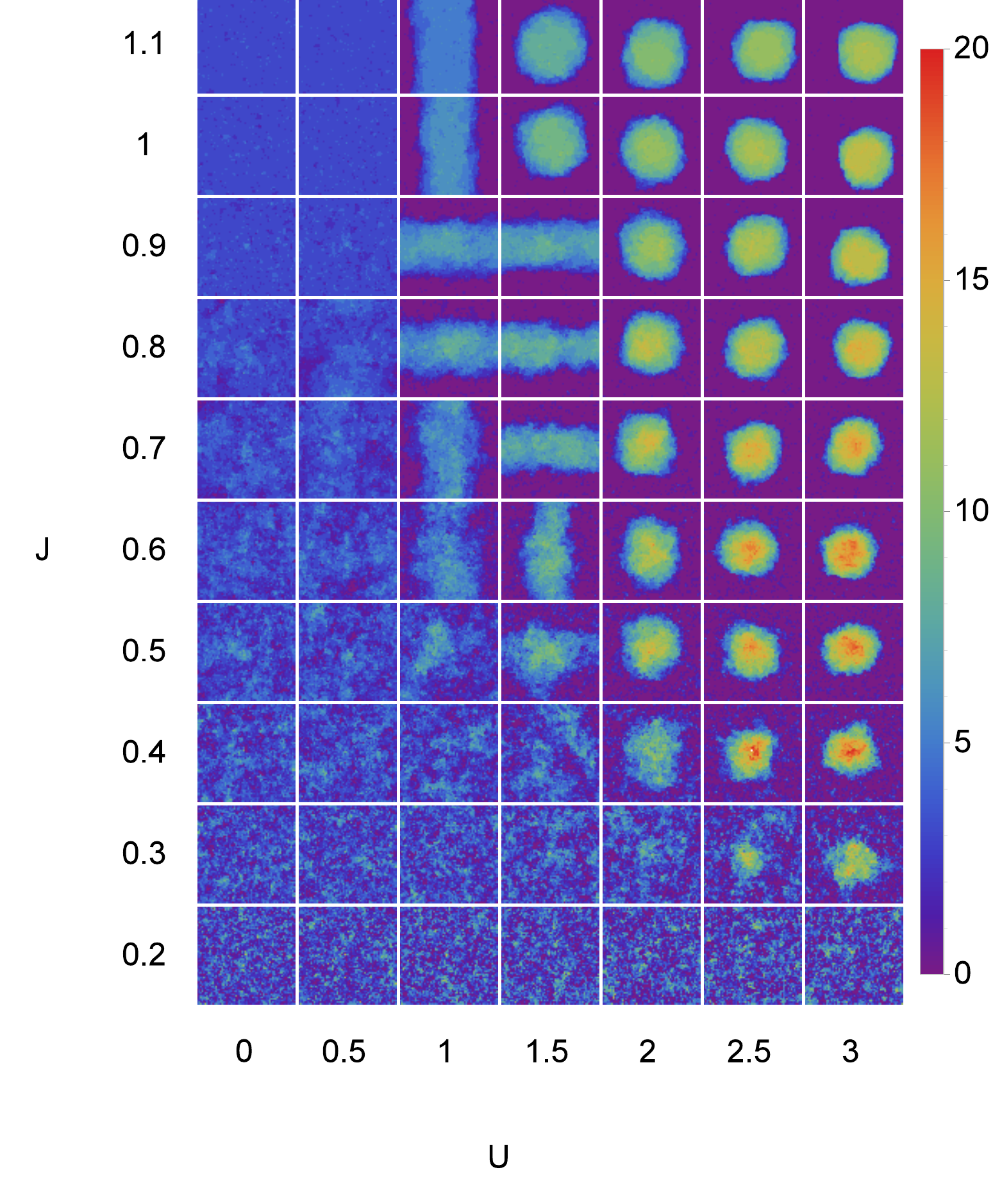}
		\includegraphics[width=0.325\textwidth]{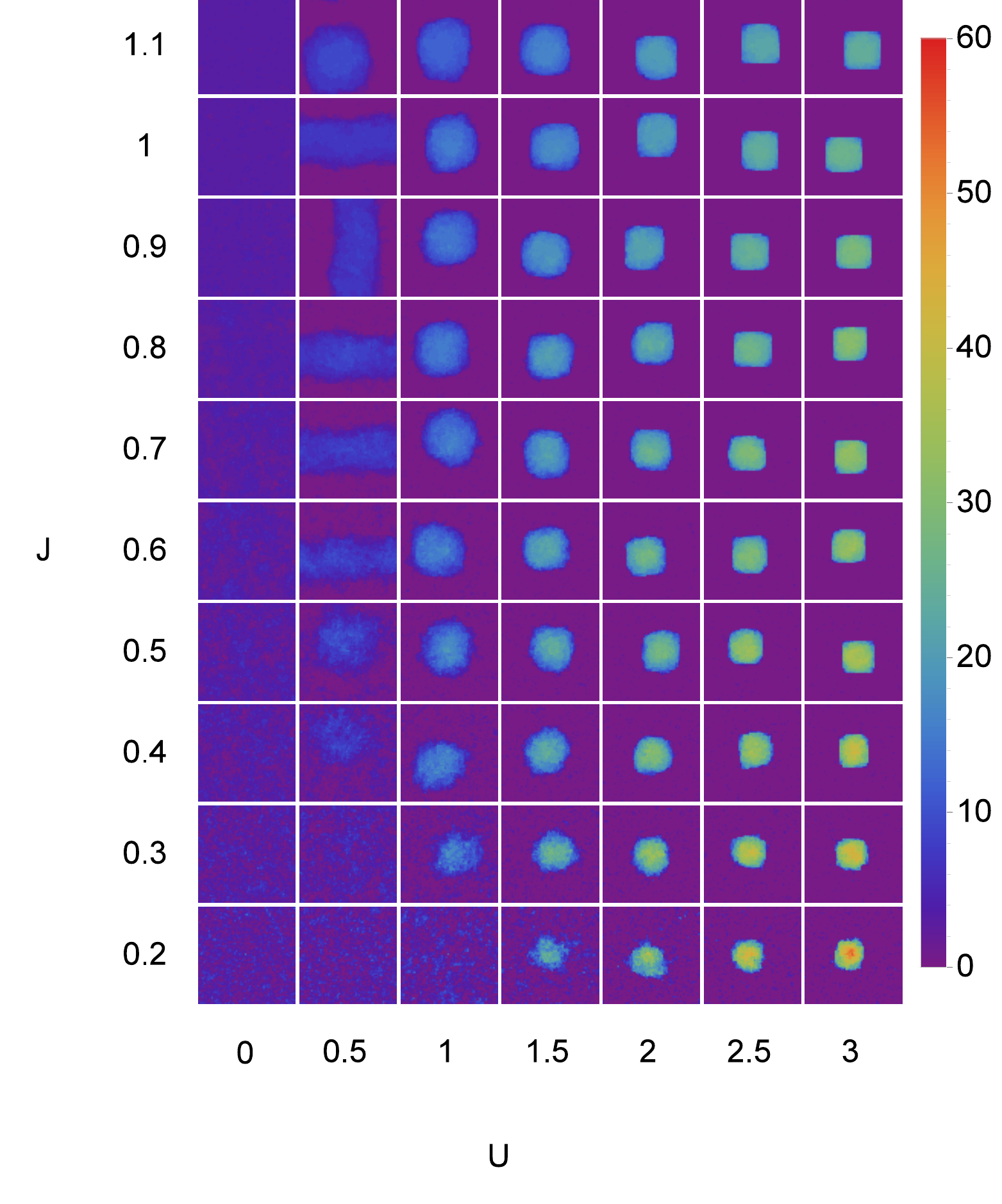}
		\includegraphics[width=0.325\textwidth]{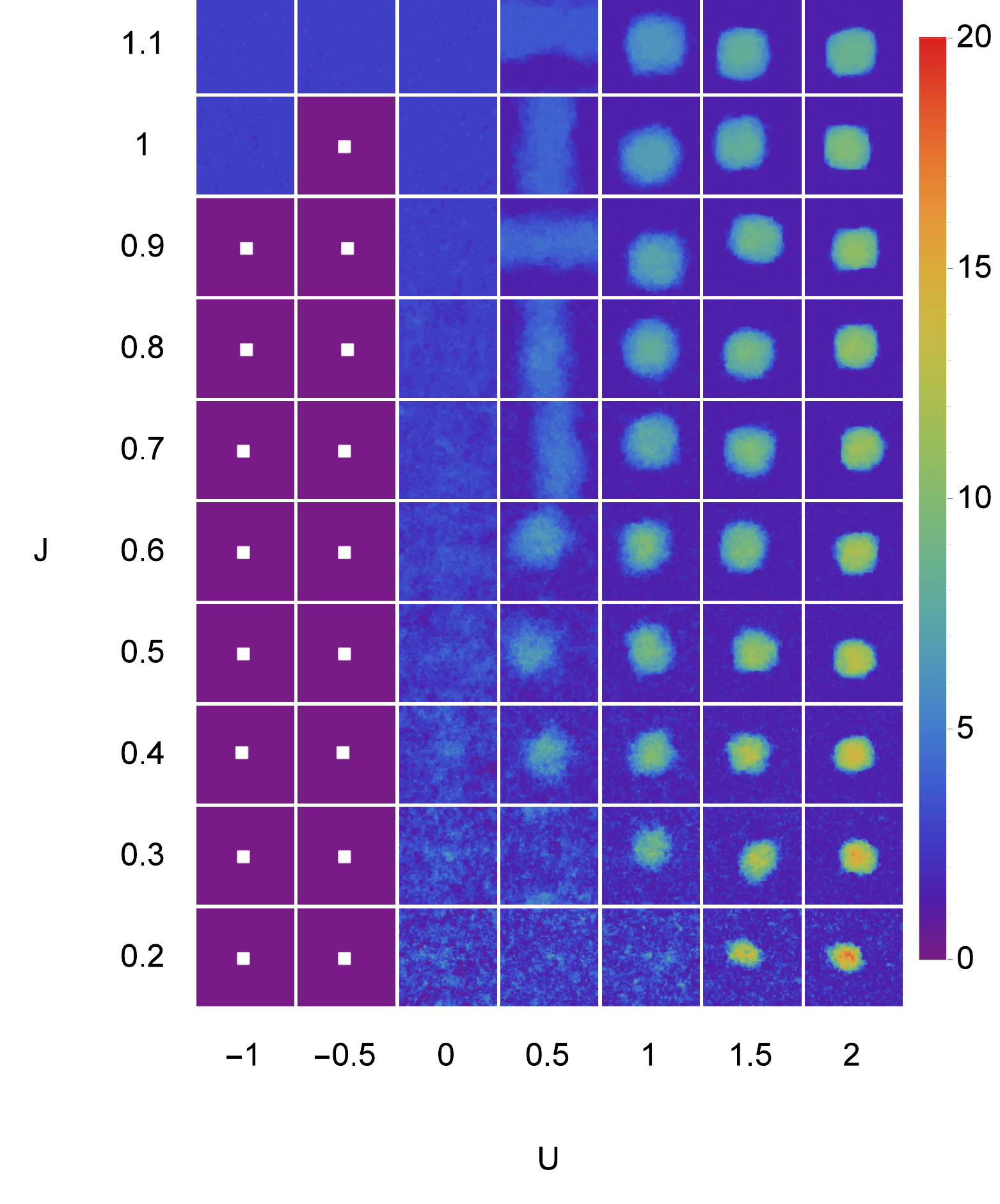}
	\caption{\label{fig:diagramS050930} (Color online) Phase diagrams for $\sigma=0.5, 0.83,$ and $3$ (left to right). Colors represent different numbers of particles per site (see the color bar).}
\end{figure}

\subsection{Phase diagram}

To explore the effect of $J,U$, and $\sigma$, we have made simulations for fixed density $\rho$ while varying $J,U,\sigma$. Figure~\ref{fig:diagramS1} shows a pictorial representation of the $U,J$-phase diagram, for $\sigma=1$. The density of particles is $\rho=3$.
The upper right corner of the diagram (large positive $U,J$) corresponds to the parameter region where particles condense into islands, and the bottom left (small $J$ and negative $U$) to the region where only a fluctuating ``wetting layer'' can be observed. At the transition region between the condensate and the wetting-layer phase vertical or horizontal ``stripes'' of particles can be seen. They are caused by periodic boundary conditions: for small $U$, due to a larger extension of the condensate its opposite sides merge together and form a stripe; a larger lattice would prevent this finite-size effect and hence the region where the stripes occur in fact belongs to the condensed phase. Stripes are also present for sufficiently negative $U$, but their origin is different: the surface now repels adatoms, which tend to cluster together.

The situation is qualitatively similar for other values of $\sigma$, see Fig.~\ref{fig:diagramS050930}. For $\sigma>1$, condensation occurs for $U>0$; the surface is covered by a ``wetting layer'' composed of $\lfloor\sigma\rfloor$ layers of particles, with the condensate on the top-most layer. For $\sigma<1$, the condensate (which again happens for $U>0$) is surrounded by empty sites and there is no wetting layer. This is due to the fact that the potential $V(m)$ has a minimum at $m=0$, and is not sufficiently deep for $m>0$ (cf. Fig.~\ref{fig:potential}(a)) for the particles to be attracted enough to the substrate. Similarly, for $\sigma>1$ and $U<0$, the surface is empty apart from a localised, very high condensate peak (white spots in Fig.~\ref{fig:diagramS050930}, right panel; for high $J$ there are no peaks because large surface tension does not let the simulation leave the flat initial condition).

\begin{figure}
	\centering
	\includegraphics[width=.49\textwidth]{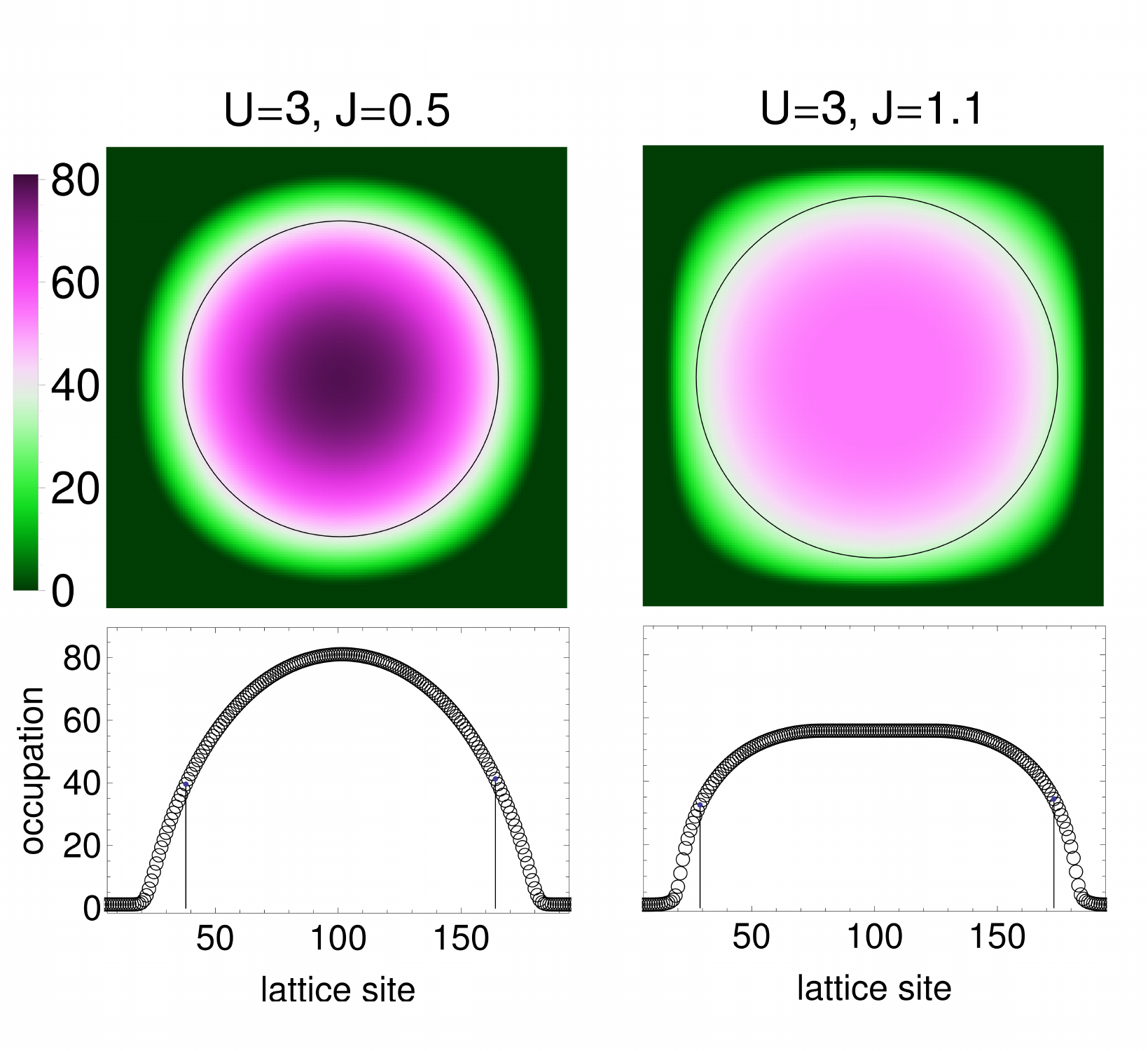}
	\caption{\label{fig:2Dshape}(Color online) Top row: condensates on a square lattice $xy$-plane) take more rectangular shapes for larger values of $J$ ($\sigma=1$, $L=200$ in both cases).
		The colour coding has been chosen so as to enhance visibility of droplet boundaries. Circles (solid lines below) correspond to half of the maximal height.
		Bottom: central cross-sections of the condensate ($xz$-plane) for the same $J,U$ as pictures in the upper row. The profile of the condensate becomes more rectangular as $J$ increases, the height decreases, and the width in the $x$-direction increases (see~\ref{app:comp}).}
\end{figure}

Figures~\ref{fig:diagramS1}--\ref{fig:diagramS050930} show that the shape of the condensate in the $xy$-plane depends on the parameters $J,\, U$: large $U$ makes the condensate narrower and higher, while large $J$ makes its surface (in the $z$-direction) flatter. Figure~\ref{fig:2Dshape}, top row, shows that on a square lattice, small $J$ leads to circularly-shaped condensates in the $xy$-plane, whereas for large $J$ the condensate assumes a more square shape, reflecting the symmetry of the underlying lattice. The same figure, bottom row, demonstrates that as the condensate becomes more rectangular, its profile (section through the centre in the $xz$-plane) changes from an approximately parabolic to a more rectangular one. We shall see in Sec.~\ref{sec:shape}, that the $xz$-profile of the (2+1)-dimensional condensate can be well approximated by the (1+1)d model.

The shape of the condensate does not depend on whether the system is in equilibrium (by making the hops symmetric: $r_k=1/4$), or not. For example, even in the extreme case when particles are allowed to jump to the right and not to the left (which produces a strong current in the $x$-direction), the steady-state shape remains unaffected. This is caused by the lack of any explicit dependence of the steady-state distribution (\ref{eq:Pm}) on the hopping probabilities $\{r_k\}$. The dynamics of condensation, however, will be different for different $\{r_k\}$.

Before we discuss the dynamics of condensation, let us briefly comment on the relation between what we see in our model, and experiments on Stranski-Krastanov growth, which partly motivated this study. As explained in the introduction, Stranski-Krastanov growth can be used to produce quantum dots. A visual comparison between our results, and experimentally obtained Al$_{x}$Ga$_{1-x}$As quantum dots on GaAs~\cite{Qdots} (Figs.~9-11 therein) and GaN on AlN~\cite{Qdots1} (Fig.~2 therein) shows that these quantum dots are not dissimilar to our condensates, and their shape depends on growth conditions and chemical composition which corresponds to different values of $J,U$ parameters in our model. Our model can therefore qualitatively reproduce certain aspects of the growth of quantum dots.

\subsection{Dynamics}
\label{sec:dynamics}

\begin{figure}
	\centering
	\includegraphics[width=.3\textwidth]{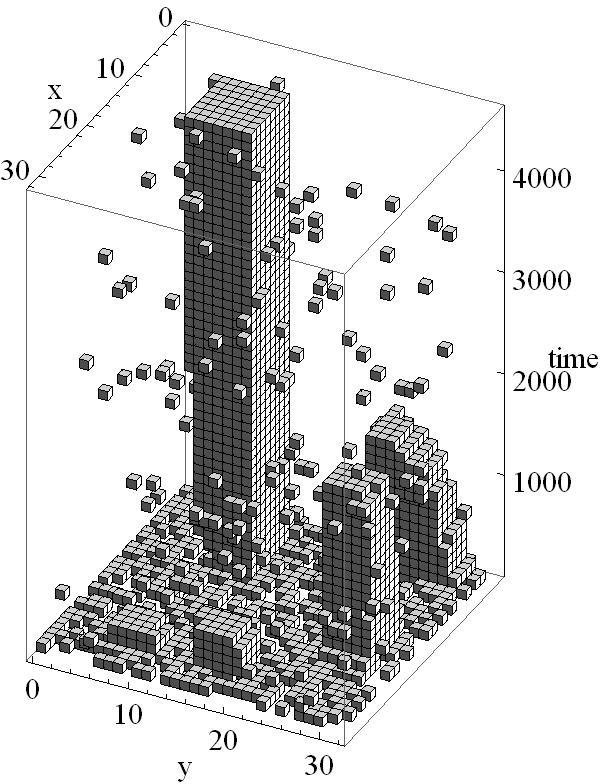}
        \hspace{2cm}
	\includegraphics[width=.3\textwidth]{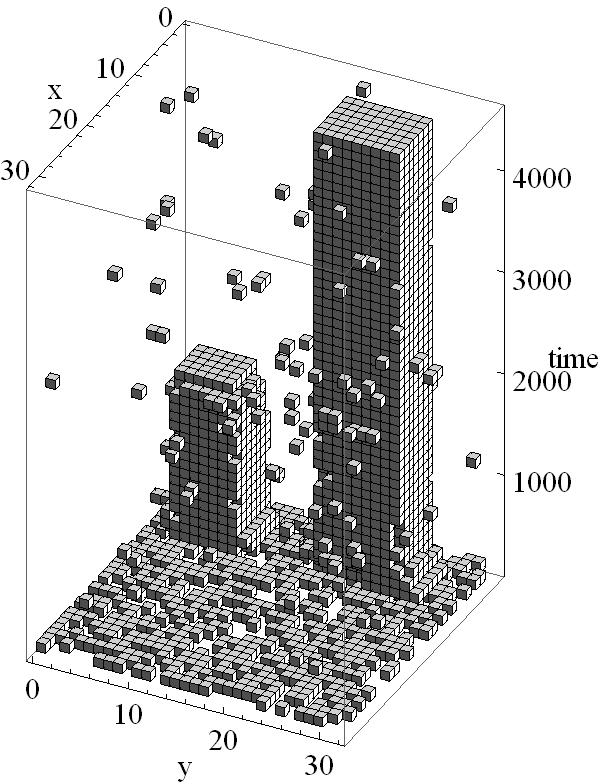}
	\caption{\label{fig:BW_kimograph}Example of the time evolution of a $32\times 32$ system, for $J=1.1, U=15, \sigma=1, \rho=2$. Each $xy$ plane (with periodic boundary conditions) corresponds to a single time frame. Cubes represent occupied sites with two or more particles (the background has 1 particle per site). Left: equilibrium model ($r_k=1/4$); right: non-equilibrium model ($r_1=r_3=r_4=1/3, r_2=0$).}
\end{figure}

We now discuss the dynamics of condensation in this model. The equilibrium Monte Carlo algorithm employed in the previous section cannot be used here and we have to simulate the process using a kinetic Monte Carlo algorithm with the hopping rate (\ref{eq:Pm}). Figure~\ref{fig:BW_kimograph}, left, shows the time evolution of the equilibrium model ($r_k=1/4$), starting from randomly distributed particles at $t=0$. The process has two phases. First, particles rapidly aggregate into clusters; the second, slower stage involves clusters exchanging particles through the background. Eventually, only one cluster -- the condensate -- remains in the system.

The time it takes for a single condensate to build up can be found using similar arguments to those for the ZRP~\cite{Evans_JPA2005}.
Numerical simulations suggest that the time $T$ to condensation is dominated by the process of merging the last few remaining clusters. Each cluster has on average $O(M)=O((\rho-\rho_{\text{c}}) L^2)$ particles and the inter-cluster distance is $O(L)$. A cluster of size $m$ loses particles through its boundary. The rate $u_{\rm emit}$ with which each of the $l=l(m)$ sites at the edge of the cluster emits particles only weakly depends on the size if $m\gg 1$. For example, the rate at which particles are emitted from a site of height $h$ in the condensate's wall is $u_{\rm emit}=\left(g(h-1,h)/g(h,h)\right)^3\left(g(h-1,\rho_{\text{c}})/g(h,\rho_{\text{c}})\right) \cong \exp(-2J)$ for any $h\gg 1$, and hence we can take it to be constant for all clusters. The total emission rate of the cluster is $lu_{\rm emit}$. Put differently, each such particle is emitted every $T_{\rm emit}=1/(lu_{\rm emit})$ time units. Once they leave the cluster, the particles undergo a random walk with diffusion constant $D\approx u(\rho_{\text{c}}+1|\rho_{\text{c}},\rho_{\text{c}},\rho_{\text{c}},\rho_{\text{c}}) = \left[g(\rho_{\text{c}},\rho_{\text{c}})/g(\rho_{\text{c}}+1,\rho_{\text{c}})\right]^4$. Most of these particles are quickly reabsorbed due to recurrence of 2d random walk \cite{Polya} but particles that have departed a distance $O(L)$ can be intercepted by other clusters. The time the particle needs to travel to reach another cluster is approximately $T_{\rm travel}=O(L^2/D)$. Since $T_{\rm travel}$ increases with $L$ whereas $T_{\rm emit}$ does not, diffusion is the limiting step and the total time it takes to move a particle from one cluster to another one is approximately equal to $T_{\rm travel}$ for large enough $L$.

\begin{figure}
	\centering
	\includegraphics[width=.45\textwidth]{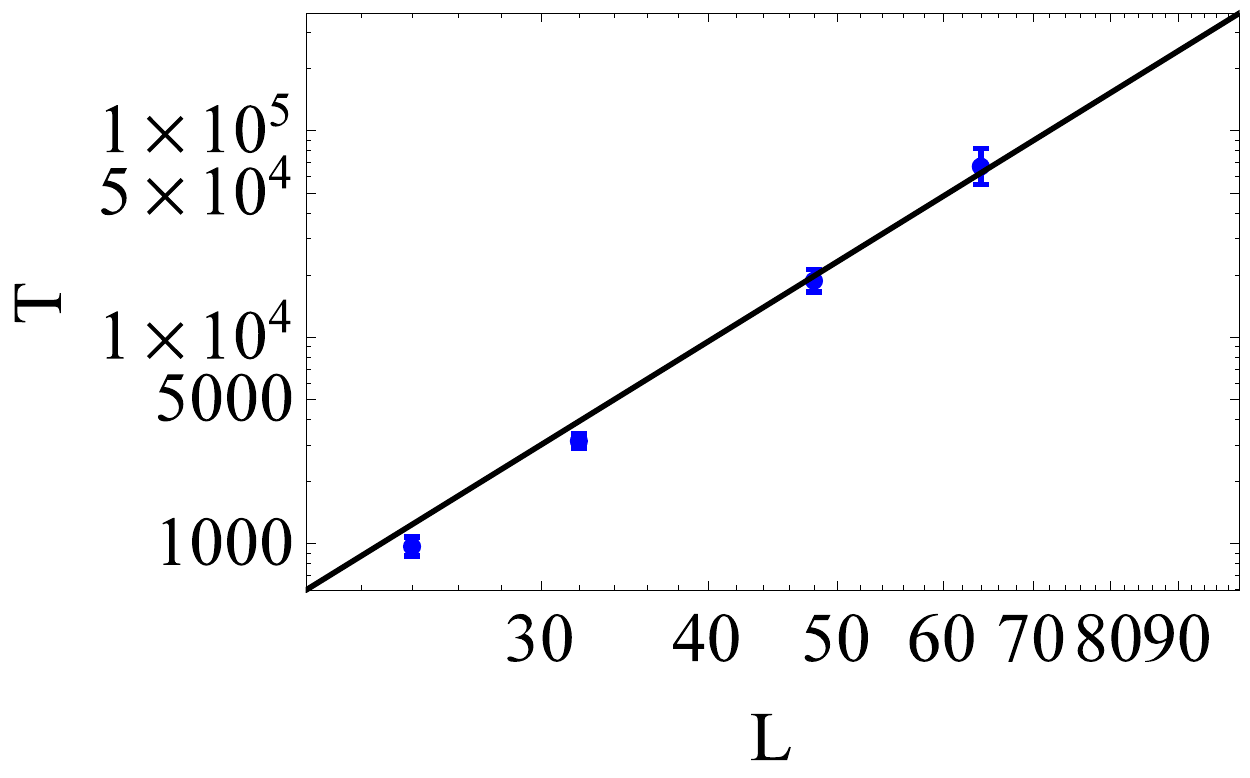}
	\includegraphics[width=.45\textwidth]{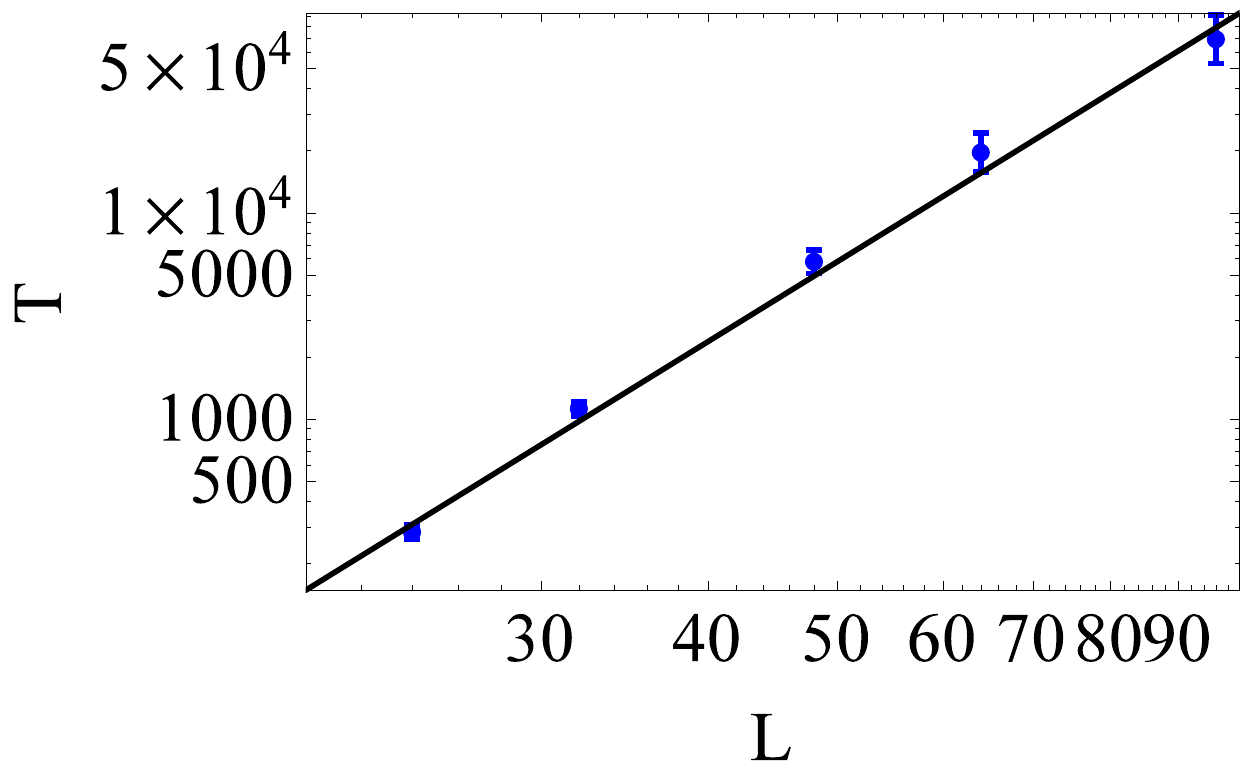}
        \caption{\label{fig:BW_Tss}The time to condensation for different sizes $L=24,32,48,64,96$. Left: equilibrium model ($r_k=1/2$), right: non-equilibrium model ($r_1=r_3=r_4=1/3, r_2=0$). In all cases $J=1.1, U=15, \sigma=1$. The solid line is $T = D^{-1} L^4$ with $D=(g(1,1)/g(2,1))^4=1076.88$ for the left panel and $D=4\times 1076.88$ for the right panel. Note that the unknown proportionality coefficient in the anticipated asymptotic formula $T\propto D^{-1}L^4$ turns out to be close to one in the non-equilibrium case and to four in the equilibrium case. }
\end{figure}

Large clusters emit more particles, but they also re-absorb them with higher probability due to their larger circumference. A particle that has diffused away is more likely to be absorbed by a large than a small cluster. This causes a net current of particles flowing from smaller to larger clusters. The time it takes to transfer $O(L^2)$ particles from a small cluster to the condensate is therefore $T=O(L^2 T_{\rm travel})=O(L^4/D)$. The scaling $T\propto D^{-1}L^4$ is verified in Fig.~\ref{fig:BW_Tss}, left, where we plot the average time it takes to have only one cluster (the condensate) in the system. 

Interestingly, the non-equilibrium case in which there is a current of particles in the $y$ direction, leads to the same prediction (Fig.~\ref{fig:BW_Tss}, right). Although individual particles drift preferably in the $y$ direction, the clusters remain quasi-static, see Fig.~\ref{fig:BW_kimograph}, right. Moreover, the time for a particle to move from one cluster to another is again $O(L^2)$; its motion is still diffusive in the $x$ direction and, unless the clusters are accidentally aligned so that particles can move between them in straight lines, diffusion dominates over the ballistic motion for which the time scale would be $O(L)$.

\section{Analytically soluble (1+1)d model}
\label{sec:1D}

Although the steady-state probability of the (2+1)d model has a simple, factorised form (\ref{eq:Pm}), its exact solution remains elusive. However, we can learn about many properties of this model using approximate methods. In this section we shall analyse a (1+1)d counterpart of the model, described by Eqs.~\eqref{eq:Um1d},\eqref{eq:Pm1d}. We shall show that this simpler model predicts not only the critical density $\rho_{\text{c}}$ but also qualitatively reproduces the transition lines in the $(U,J)$-phase diagrams from Figs.~\ref{fig:diagramS1}, \ref{fig:diagramS050930}. 

The model can be analysed along the same lines as in Ref.~\cite{BW_JStat}. Let us define the canonical partition function
\beq
Z(L,M)=\sum_{\{ m_i \}}\prod_{i=1}^{L}{g(m_i,m_{i+1})\delta\left[\sum_{i=1}^{L}{m_i}-M\right]},
\eeq
and its grand-canonical counterpart
\beq
Z_L(z)= \sum_{M}Z(L,M)z^M = \sum_{\{ m_i \}}z^{\sum_{i=1}^{L}{m_i}}\prod_{i=1}^{L}{g(m_i,m_{i+1})},
\eeq
where $z$ is the fugacity, determined from its relationship to the average density,
\beq
\label{eq:rho}
\rho(z)=\frac{1}{L}\left\langle \sum_{i}{m_i}\right\rangle = \frac{z}{L}\frac{\partial\ln Z_L(z)}{\partial z}.
\eeq	
Thanks to the one-dimensionality of the problem, $Z_L(z)$ can be expressed using the standard transfer-matrix approach:
\beq
Z_L(z)= \sum_{m_1,\ldots,m_L}{T_{m_1 m_2}T_{m_2 m_3}\cdots T_{m_L m_1}} = \mbox{Tr}T(z)^L,
\eeq
where $T_{mn}=z^{(m+n)/2}g(m,n)$ denotes the transfer matrix. We expect the partition function $Z_L(z)$ to have a finite radius of convergence $z_{\text{c}}$. If $\rho(z)\to\infty$ as $z$ increases from $0$ to $z_{\text{c}}$, the grand-canonical ensemble is valid for any density of particles, and the probability $p(m)$ of finding $m$ particles at a randomly chosen site reads
\beq
	p(m) = \frac{\phi_m^2}{\sum_{m=0}^{\infty} \phi_m^2},
\eeq
where $\phi_m$ is the principal eigenvector (associated with the largest eigenvalue) of $T_{mn}(z)$. If, however, $\rho(z)$ has a finite limiting value $\rho(z) \longrightarrow \rho_{\text{c}}$ as $z$ approaches $z_{\text{c}}$, then Eq.~\eqref{eq:rho} cannot be satisfied for $z\geq z_{\text{c}}$ (or equivalently for $\rho>\rho_{\text{c}}$) and the grand-canonical ensemble cannot be constructed. This corresponds to a transition from the liquid to the condensed state for $\rho>\rho_{\text{c}}$.

Since for our choice (\ref{eq:gmndef}) of $g(m,n)$ the critical $z_{\text{c}}=1$, to determine the critical density at which this transition happens, we must find the eigenvector $\phi_m$ of the matrix $T_{m,n}=g(m,n)$ to the maximal eigenvalue $\lambda_{\rm{max}}$:
\beq
\label{eq:eigenproblem}
\sum_n g(m,n) \phi_n = \lambda_{\rm{max}} \phi_m,
\eeq
and, rewriting the partition function in the large-$L$ limit as $Z_L(z)\cong \lambda^L_{\rm{max}}$, from Eq.~(\ref{eq:rho}) we obtain
\beq
\rho_{\text{c}}=\frac{\sum_{m=0}^\infty{m \phi_m^2}}{\sum_{m=0}^{\infty} \phi_m^2}. \label{eq:rhocdef}
\eeq
Condensation can occur only if $\phi_m$ decays with $m$ faster than $\sim m^{-1}$, otherwise the critical density $\rho_{\text{c}}$ becomes infinite in the thermodynamic limit. It turns out that although the eigenproblem (\ref{eq:eigenproblem}) is very easy to solve numerically, it is still too difficult to solve analytically for our particular choice of $g(m,n)$ from Eq.~(\ref{eq:gmndef}). To make progress, we observe that since the occupation numbers are discrete, the potential values are discretised as well. Moreover, the value of $V(m)$ varies significantly with $m$ only for the first few integer $m$. This suggests that the potential can be approximated by a sum of a few Kronecker delta functions with appropriate amplitudes.

\subsection{Solution for $\sigma\ll 1$}
\label{sec:modelA}
 
Let us first consider the case $\sigma\ll 1$. As illustrated in Fig.~\ref{fig:potential}(a), the on-site potential $V(m)$ has one dominant minimum. This allows us to approximate the potential as $-\tilde{U} \delta\left[ m \right]$, where $\delta\left[ m \right]$ is the Kronecker delta, and $\tilde{U}=U (\sigma^3 - \sigma^9)$. The value of $\tilde{U}$ is chosen to reproduce the value obtained from the exact formula \eqref{eq:LJmodel} for $m=0$. Figure~\ref{fig:potential}(b) shows this approximate potential for $\sigma=0.5$. The term $\delta\left[ m \right]$ lowers the energy and hence it increases the probability of a state in which the occupation is $m=0$; non-zero occupation, on the other hand, is energetically unfavourable. Physically, this could mean that the particles cannot wet the substrate that is strongly ``hydrophobic''. Therefore, the on-site potential favours empty sites, which leads to mass condensation seen as an ``island'' of particles discussed before.

This model, which we shall call ``model A'' here, can be solved using the same approach as in Ref.~\cite{BW_JStat}. Assuming the weight matrix 
\beq
\label{eq:gmnDelta}
g(m,n)=\exp[-J|m-n|+\tilde{U}(\delta\left[ m \right]+\delta\left[ n \right])/2],
\eeq
the eigenvector $\phi_m$ which solves Eq.~\eqref{eq:eigenproblem} must take the form $\phi_m \propto \exp(A\delta\left[ m \right]+B m)$, with some constants $A,B$. Inserting it into Eq.~\eqref{eq:eigenproblem} we obtain the constants $A=\tilde{U}/2, B=-J-\ln[1-\exp(-\tilde{U})]\equiv-J+J_0$, where $J_0=-\ln[1-\exp(-\tilde{U})]$. Since $\phi_m\propto \exp(B m)$, the entries of $\phi_m$ increase with increasing $m$ for $J<J_0$ and, from Eq.~(\ref{eq:rhocdef}), the critical density $\rho_{\text{c}}$ is infinite. This means that condensation cannot occur even for a very high density $\rho$ of particles if $J<J_0$. For $J>J_0$, however, $\phi_m$ falls off exponentially, and the critical density for model A reads
\beq
\rho_{\text{c}}^{A}=\frac{\sum_{m=0}^{\infty} m \phi_m^2}{\sum_{m=0}^{\infty} \phi_m^2} = \frac{e^{J_0}-1}{\left[e^{J_0}-e^{-2 (J-J_0)}\right]\left[e^{2 (J-J_0)}-1\right]}.
\label{eq:rhocA}
\eeq
For example, for $\sigma=0.5$ and $U=J=2$ (where $J>J_0\approx 1.5$), the critical density $\rho_{\text{c}}\approx 0.53$.
The transition line, which separates the region in the $(U,J)$-plane where condensation occurs from the region where it does not, is given by
\beq
\label{eq:transLine}
J=-\ln(1-e^{-U(\sigma^3 - \sigma^9)}).
\eeq
In Fig.~\ref{fig:comparison1D} we show a plot of Eq.~(\ref{eq:transLine}) compared with an exact, numerical solution to the eigenproblem \eqref{eq:eigenproblem}. The agreement is good even for a relatively large $\sigma=0.5$. It is also worth noticing that the transition line predicted by this model is qualitatively similar to that of the (2+1)d model from Fig.~\ref{fig:diagramS050930}.

\subsection{Solution for $\sigma\gg 1$}
\label{sec:modelB}

\begin{figure}
	\centering
		\includegraphics[width=.49\textwidth]{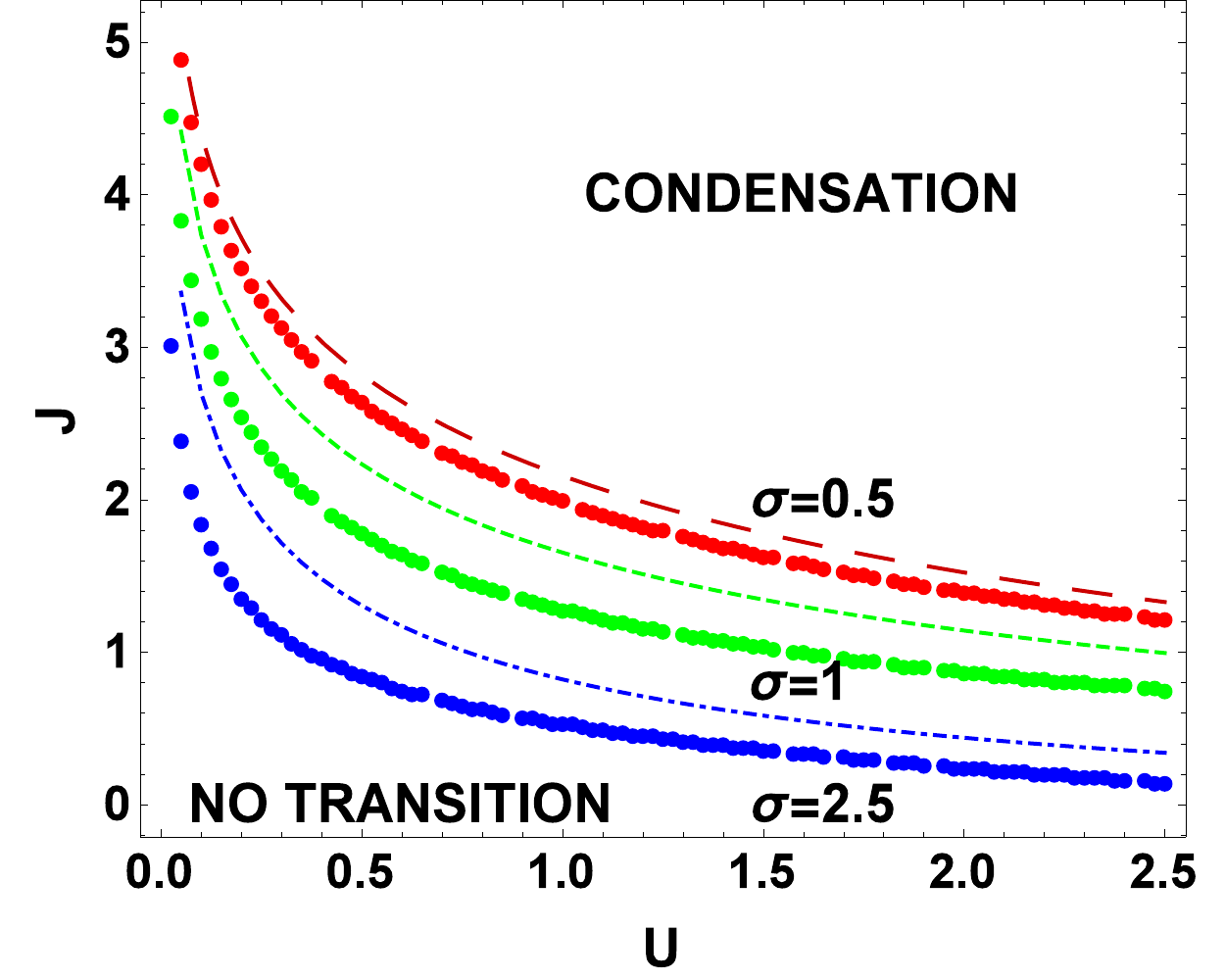}
		\caption{\label{fig:comparison1D}(Colour online)
		Phase transition lines for the (1+1)d model with LJ potential \eqref{eq:LJmodel}.
		The dashed lines are analytic solutions: The long-dashed curve corresponds to model A [Eq.~\eqref{eq:transLine}], the short-dashed and the dot-dashed to the double-delta model C [Eq.~\eqref{eq:transLineC}].
		The points are numerical solutions obtained by diagonalising $g(m,n)$ numerically (see~\ref{app:rhoc}).
		}
\end{figure}

In the case $\sigma\gg 1$, the minimum of the potential is located at $m=\left\lfloor \sigma\right\rfloor$. Since for all occupation numbers $m< \left\lfloor \sigma\right\rfloor$ the potential is very large (see Fig.~\ref{fig:potential}), we can approximate it by
\beq
\label{eq:Bpm}
V(m)\approx \left\{ \begin{array}{ll} -\tilde{U}\delta\left[ m-\left\lfloor \sigma\right\rfloor\right], & \mbox{for}\; m\geq \left\lfloor \sigma\right\rfloor, \\ \infty, & \mbox{for}\; m<\left\lfloor \sigma\right\rfloor, \end{array} \right.
\eeq
where
\beq
\label{eq:BUtilde}
\tilde{U}=U \left[\left(\frac{\sigma}{\left\lfloor \sigma\right\rfloor+1}\right)^3 - \left(\frac{\sigma}{\left\lfloor \sigma\right\rfloor+1}\right)^9\right].
\eeq
Let us call this ``model B''. The potential barrier at $m_0=\left\lfloor \sigma\right\rfloor -1$ means that if the density of particles $\rho>\left\lfloor \sigma\right\rfloor$, no site will have less than $\left\lfloor \sigma\right\rfloor$ particles. Consequently, all elements $m=0,\dots,\left\lfloor \sigma\right\rfloor-1$ of the eigenvector $\phi_m$ will be zero. We can derive the critical density for condensation in the same way as in the previous section. Assuming that $\phi_{m} \propto \exp(A\delta\left[m-\left\lfloor \sigma\right\rfloor\right]+B m)$ for $m\geq \left\lfloor \sigma\right\rfloor$, we obtain that $A=\tilde{U}/2, B=-J+J_0$, where $J_0=-\ln[1-\exp(-\tilde{U})]$ has the same form as previously, albeit with a different $\tilde{U}$ given by Eq.~(\ref{eq:BUtilde}). The critical density given by Eq.~(\ref{eq:rhocA}) is shifted by $\left\lfloor \sigma\right\rfloor$ which accounts for the vanishing elements of the eigenvector (the shift by $x$ occurs for any $g(m,n)$ whose eigenvector's first $x$ elements vanish):
\beq
\label{eq:linDens2}
\rho_{\text{c}}^{B}= \left\lfloor \sigma\right\rfloor + \rho_{\text{c}}^{A} = \left\lfloor \sigma\right\rfloor + \frac{e^{J_0}-1}{\left[e^{J_0}-e^{-2 (J-J_0)}\right]\left[e^{2 (J-J_0)}-1\right]}.
\eeq
The critical line $J(U)$ separating the condensed and liquid phases is
\beq
\label{eq:transLineB}
J=-\ln\left\{1-\exp\left[-U \left(\left(\frac{\sigma}{\left\lfloor \sigma\right\rfloor+1}\right)^3 - \left(\frac{\sigma}{\left\lfloor \sigma\right\rfloor+1}\right)^9\right) \right]\right\}.
\eeq
This expression is a good approximation for the critical line for large $\sigma$, but is much worse for $\sigma\approx 2-3$ that we use here in simulations. In the next section, \ref{sec:modelC}, we show that if the potential $V(m)$ is approximated by a sum of two delta functions (``model C''), the agreement between the approximate solution $J(U)$ [see Eq.~(\ref{eq:transLineC})] and the simulation data becomes much better for smaller $\sigma$, as seen in Fig.~\ref{fig:comparison1D}. 

\subsection{Double-delta approximation of the potential}
\label{sec:modelC}
 
In previous sections we used a single-Kronecker delta approximation of the LJ potential. This approximation is fairly sufficient for the two limits from Secs.~\ref{sec:modelA},B for $\sigma\ll 1$ and $\sigma\gg 1$, but it does not work well for $\sigma\approx 1$ which is the range we are interested in in this work. The approximation can be improved by modelling the potential with two Kronecker delta functions with suitable amplitudes:
\beq
\label{eq:Dpm}
V(m)= \left\{ \begin{array}{ll} -\tilde{U}_1\delta \left[m-\left\lfloor \sigma\right\rfloor\right]-\tilde{U}_2\delta\left[ m-\left\lfloor \sigma\right\rfloor -1\right], & \mbox{for}\; m\geq \left\lfloor \sigma\right\rfloor \\ \infty, & \mbox{for}\; m<\left\lfloor \sigma\right\rfloor \end{array} \right.
\eeq
where
\beq
\label{eq:DUtilde}
\tilde{U}_1=U \left[\left(\frac{\sigma}{\left\lfloor \sigma\right\rfloor+1}\right)^3 - \left(\frac{\sigma}{\left\lfloor \sigma\right\rfloor+1}\right)^9\right] \mbox{ and} \quad
\tilde{U}_2=U \left[\left(\frac{\sigma}{\left\lfloor \sigma\right\rfloor+2}\right)^3 - \left(\frac{\sigma}{\left\lfloor \sigma\right\rfloor+2}\right)^9\right].
\eeq
The potential (\ref{eq:Dpm}) is shown in Fig.~\ref{fig:potential}(b) for $\sigma=1$ and $2.5$ (blue continuous lines). We shall refer to this model as ``model C''. The principal eigenvector of $g(m,n)$ for this model is given by $\phi_{n} \propto \exp(A_1\delta\left[ n-\left\lfloor \sigma\right\rfloor\right]+A_2\delta\left[ n-\left\lfloor \sigma\right\rfloor-1\right]+B n)$ (and $\phi_0=\ldots=\phi_{\left\lfloor \sigma\right\rfloor-1}=0$), where $A_2=\tilde{U}_2/2$. The parameters $A_1$ and $B$ can be determined analytically from the third order polynomial equation in $e^B$
\begin{align}
\label{eq:DDAB}
0 &= e^{A_{1}+U_{1}/2}+e^{B+J+U_2}-\frac{e^{2 (B+J)}}{e^{B+J}-1},
\end{align}
on substitution of $A_1$
\begin{equation}
A_1 = \frac{U_{1}}{2}+2B-\ln \left[ e^{B+J}+e^{U_1} \left( 1 -2 e^B \cosh J + e^{2B} \right) \right].
\end{equation}
The transition line, which corresponds to the condition $B=0$, can be determined from a cubic equation in $e^J$,
\beq
\label{eq:transLineC}
1-e^{-J}=\left[ e^{U_2} \left(1-e^J\right)+e^J\right] \left(2-2 \cosh J + e^{J-U_1}\right). 
\eeq
This equation can be solved exactly, but the formula for $J=J(U)$ is complicated and not very illuminating, hence we omit it here and only plot the solution in Fig.~\ref{fig:comparison1D}.

\subsection{Shape of the condensate}
\label{sec:shape}

\begin{figure}
	\centering
		\includegraphics[width=.49\textwidth]{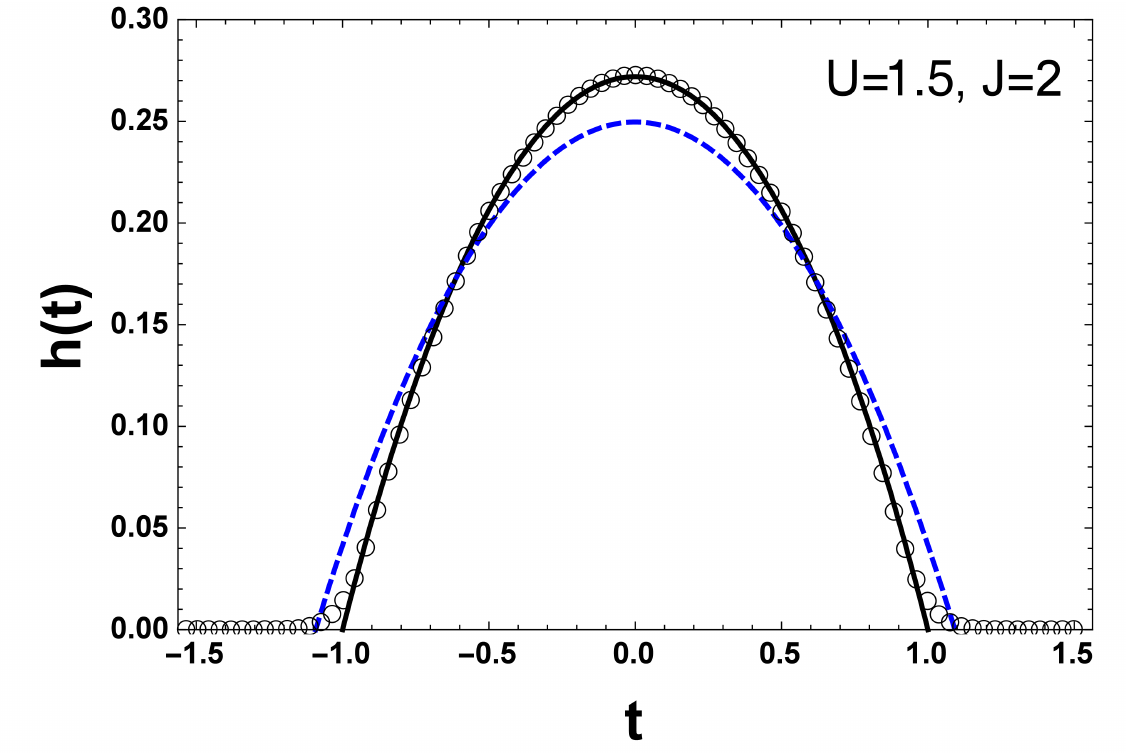}
		\hfill
		\includegraphics[width=.49\textwidth]{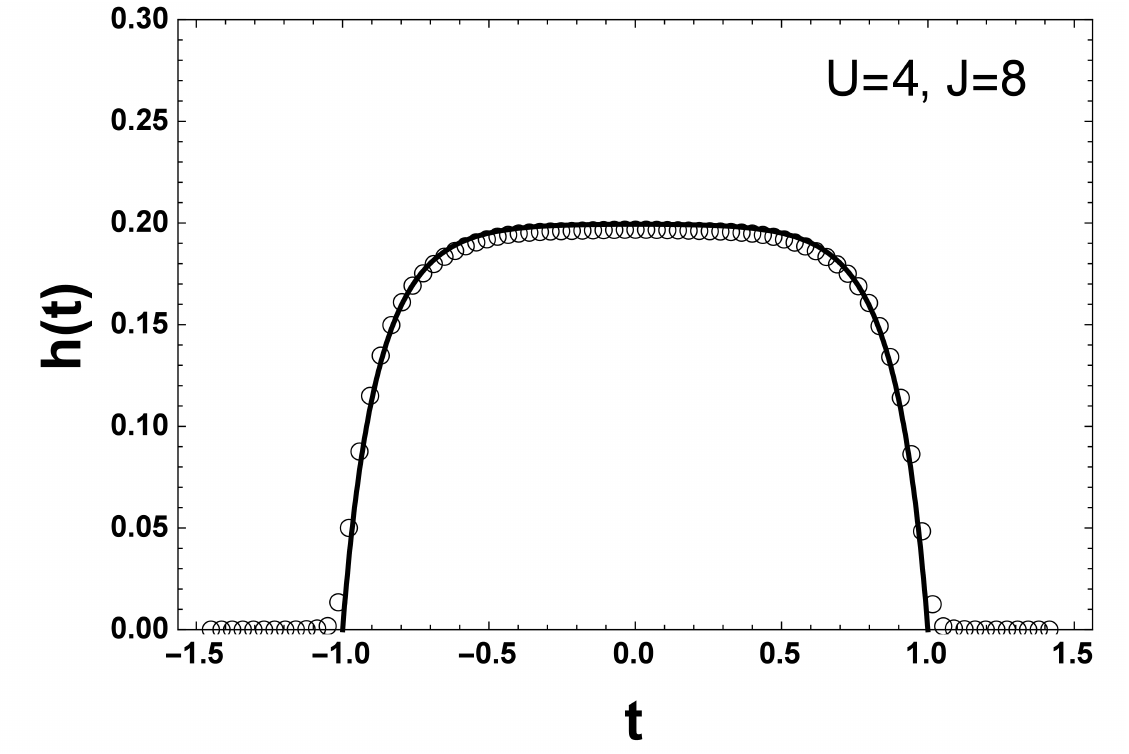}
		\caption{\label{fig:1Dshape} The shape of the (1+1)d condensate plotted in the normalised variables $(t,h)$ for $\sigma=1$. The curves represent Eq.~(\ref{eq:shape}), where for black continuous lines $\lambda_{\rm max}$ was obtained by numerical diagonalisation of a $100 \times 100$ $g(m,n)$ matrix, and for blue dashed line by solving model C ( the two curves are identical in the right panel); the circles come from a MC simulation of LJ system of size $L=2000$ with $M=60 000$ particles, averaged over $10^7$ MC sweeps. The layer of thickness $\rho_{\text{c}} \approx 1.6$ and $1.0$ has been subtracted in the left and right panel, respectively. The actual height and width of the condensate are $h(0)\sqrt{M'}\approx 65$ and $w\sqrt{M'}\approx 1302$ for the left panel, and $h(0)\sqrt{M'}\approx 48$ and $w\sqrt{M'}\approx 1379$ for the right one, respectively.
		}
\end{figure}

Above $\rho_{\text{c}}$, a spatially extended condensate forms in the system, see Fig.~\ref{fig:1Dshape}. The figure shows that the shape of the condensate, obtained from MC simulations by shifting the condensate to $i=L/2$ and averaging over many samples, is approximately parabolic. The shape can be analytically derived using the result of Ref.~\cite{BW_JStat} for a (1+1)d model with a weight function $g(m,n)=K(|m-n|)\sqrt{p(m)p(n)}$, where $K(m)$ and $p(m)$ are arbitrary functions decaying sufficiently fast with $m	\to\infty$. Let us assume that the condensate has mass $M'=M-\rho_{\text{c}} L$, and define rescaled variables $h\equiv \langle m_i\rangle/\sqrt{M'}$ and $t=\frac{2i}{w \sqrt{M'}}-1$, where $\langle m_i\rangle$ denotes the mean occupation of lattice site $i$ and $w$ is a constant which we shall determine later. In the large-$L$ limit, the shape of the condensate in these new variables is given by~\cite{BW_JStat}
\beq
\label{eq:shape0}
h(t)=\frac{w}{2v}\ln \frac{\tilde{K}(v)}{\tilde{K}(vt)} ,
\eeq
where $\tilde{K}(x)\equiv \sum_{m=-\infty}^{+\infty}K(|m|)e^{m x}$, and $w$ and $v$ are auxiliary parameters that can be obtained from $\lambda_{\rm{max}}=\tilde{K}(v)$ and 
\beq
 w=v \left(\frac{1}{2} \int^{v}_{0}{\frac{x \tilde{K}'(x)}{\tilde{K}(x)}dx}\right)^{-1/2}. \label{eq:wv}
\eeq
The one-point weight function $p(m)$ enters these formulas only through the largest eigenvalue $\lambda_{\rm max}$ of the matrix $g(m,n)$. Applying these results to our model with LJ potential, we have $K(m)=e^{-Jm}$, and hence the function $\tilde{K}(x)=\frac{\sinh J}{\cosh J - \cosh x}$. The shape $h(t)$ reads
\beq
\label{eq:shape}
h(t)=\frac{w}{2v}\ln \left(\frac{\cosh J- \cosh vt}{\cosh J - \cosh v}\right),
\eeq
where $v$ must be determined from the equation
\beq
\label{eq:lambda}
	\lambda_{\rm{max}}= \frac{\sinh J}{\cosh J - \cosh v}.
\eeq
As already mentioned, the only dependence on the potential $V(m)=-\ln p(m)$ is through the eigenvalue $\lambda_{\rm max}$ of $g(m,n)$, which can be found numerically for the LJ potential, and analytically for the approximate models A--C, for which
\beq
	v = J - J_0, 
\eeq
and $J_0$ is obtained from Eqs.~(\ref{eq:transLine}), (\ref{eq:transLineB}), and (\ref{eq:transLineC}), for the respective models.
Equation (\ref{eq:shape}) is a good approximation to the exact shape of the (1+1)d condensate already for relatively small $M$, see Fig.~\ref{fig:1Dshape}.

\begin{figure}
	\centering
		\includegraphics[width=.49\textwidth]{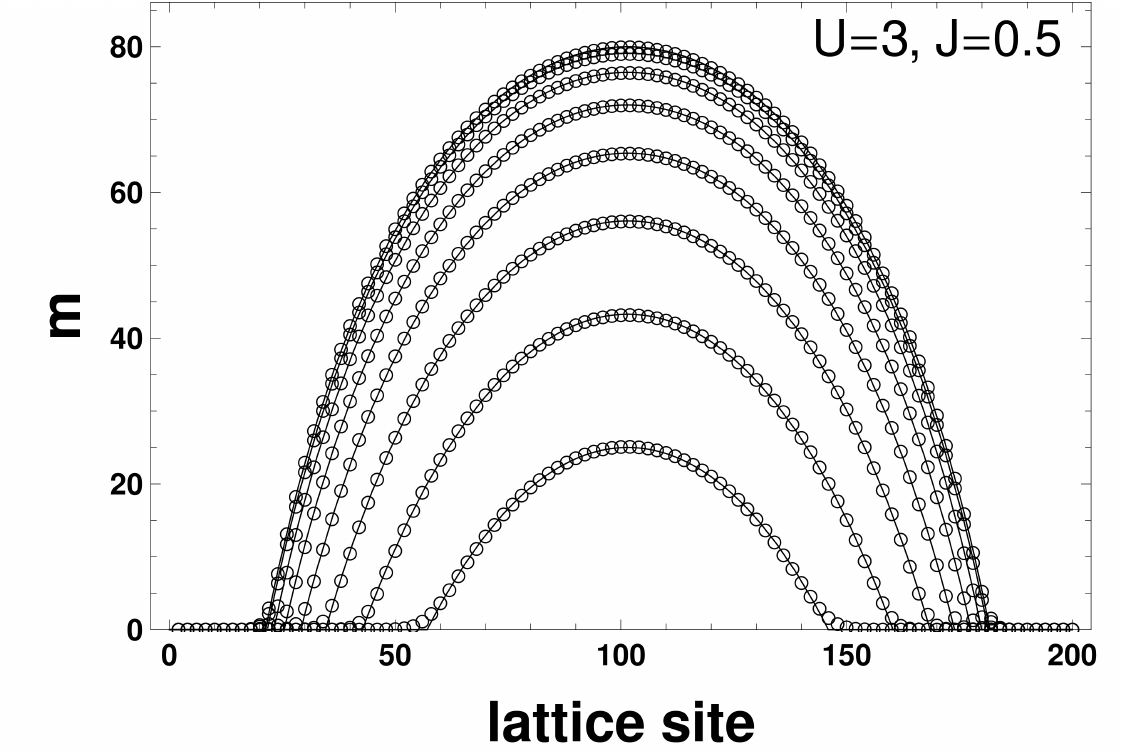}
		\hfill
		\includegraphics[width=.49\textwidth]{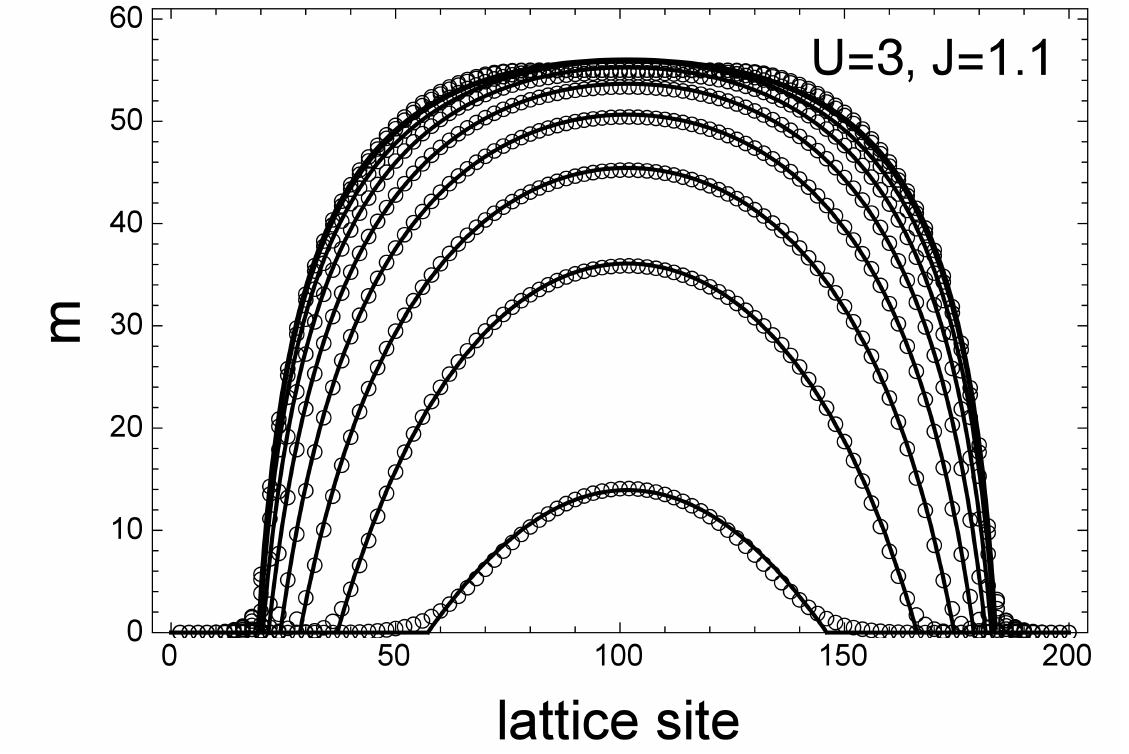}
		\caption{\label{fig:secSide}
		Cross-sections of the (2+1)d condensate with LJ potential ($\sigma=1$) in $xz$ plane (circles). The biggest envelope corresponds to the most-central section ($y=0$), with other sections taken at $y=10, 20, 30, \dots$ lattice sites. In sum, $8$ sections are plotted in the left and $9$ in the right panel. The continuous lines are the (1+1)d theoretical shapes \eqref{eq:shape} fitted with only two parameters $v,J$, and $w=w(v,J)$ obtained from Eq.~\eqref{eq:wv}. The parameters $v,J$ were fitted separately for each section; the ``effective'' parameters $J$ and $U=-\ln \left[(1+1)/(e^{J-v}-1) \right]$ (an exact solution for the (1+1)d delta model) obtained from the fit decrease monotonously with increasing distance $y$ from the central section. The details of the simulation can be found in~\ref{app:comp}.
		}
\end{figure} 

It is also interesting to note that the two-dimensional sections of the (2+1)d condensate resemble very closely the (1+1)d envelope. In Fig.~\ref{fig:secSide} we show that the $xz$-section through the centre of the condensate as well as further non-central sections are very well approximated by Eq.~(\ref{eq:shape}), with $v,J$ fitted to the numerical data. However, we do not know whether this similarity is not a mere coincidence, nor could we find an analytical formula which would predict the ``effective'' constants $w,v$ from the ``bare'' parameters $U,J$ of the (2+1)d model.

\section{Deposition of new particles with constant rate}
\label{sec:influx}

One of the features of our non-equilibrium model is that its steady-state probability assumes a relatively simple, factorised form (\ref{eq:Pm}) and, as we have seen, this allows us to calculate some quantities analytically. In this section, we explore the consequences of breaking this factorisation by releasing the constraint of mass conservation.

In the new model, particles are added to the system at a constant rate $\alpha$, as in molecular beam epitaxy. This model does not have a steady state in the sense of the constant-mass model from previous sections, because the number of particles per site increases over time. However, we shall see that the model has a quasi-steady state when the number of deposited particles is not too large, and that this state is very similar to what we discussed before.

\begin{figure}
	\centering
	\includegraphics[width=\textwidth]{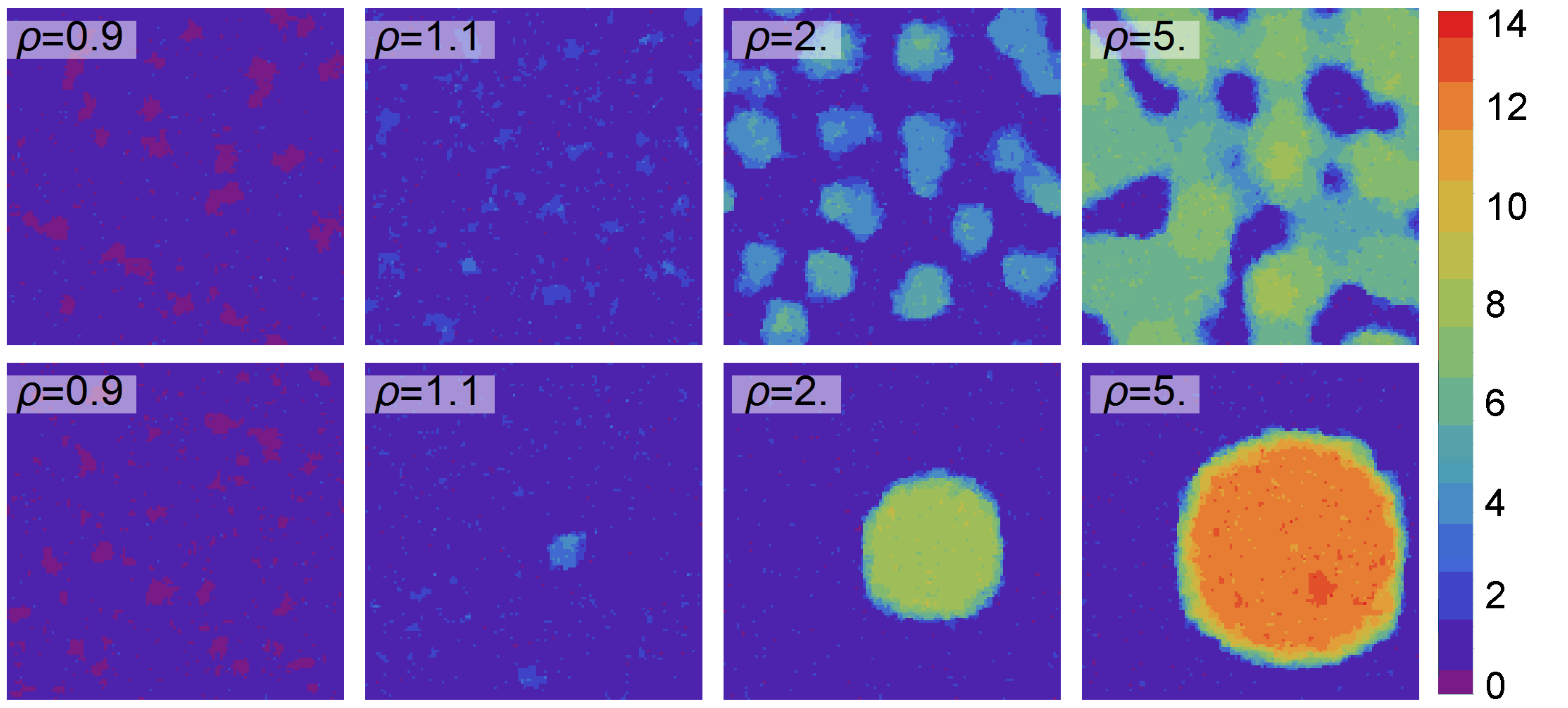}
		\caption{\label{fig:dynamic2D}(Colour online)
Simulations of a non-equilibrium system with $J=1.1, U=3, \sigma=1$ on a $128 \times 128$ lattice for a constant rate of mass deposition: (top row) $\alpha=9.85$ incoming particles per unit time, (bottom row) $\alpha=0.31$ particles per unit time. The time ranges (top) from $1.5 \times 10^{3}, 1.8 \times 10^{3}, 3.3 \times 10^{3}$ to $8.3 \times 10^{3}$ and (bottom) from $47.9 \times 10^{3}, 58.5 \times 10^{3}, 106.4 \times 10^{3}$ to $266 \times 10^{3}$ time units. Multiple condensates  form when the deposition rate is high enough (top), whereas for low $\alpha$ (bottom) only one condensate is created.
		}
\end{figure}

Figure~\ref{fig:dynamic2D} shows snapshots of the system at different times, for two different (low and high) mass deposition rates. For low deposition rates it can be seen that a single condensate is formed. This is not unexpected: particles jump between lattice sites much faster than it takes to add a new particle, and the system relaxes to a quasi-steady state similar to that of the constant-mass model. However, when the deposition rate $\alpha$ is high enough, new condensates are formed faster than they can coalesce. In this regime multiple condensates arise. 

\begin{figure}[htbp]
	\centering
	\includegraphics[width=0.49\textwidth]{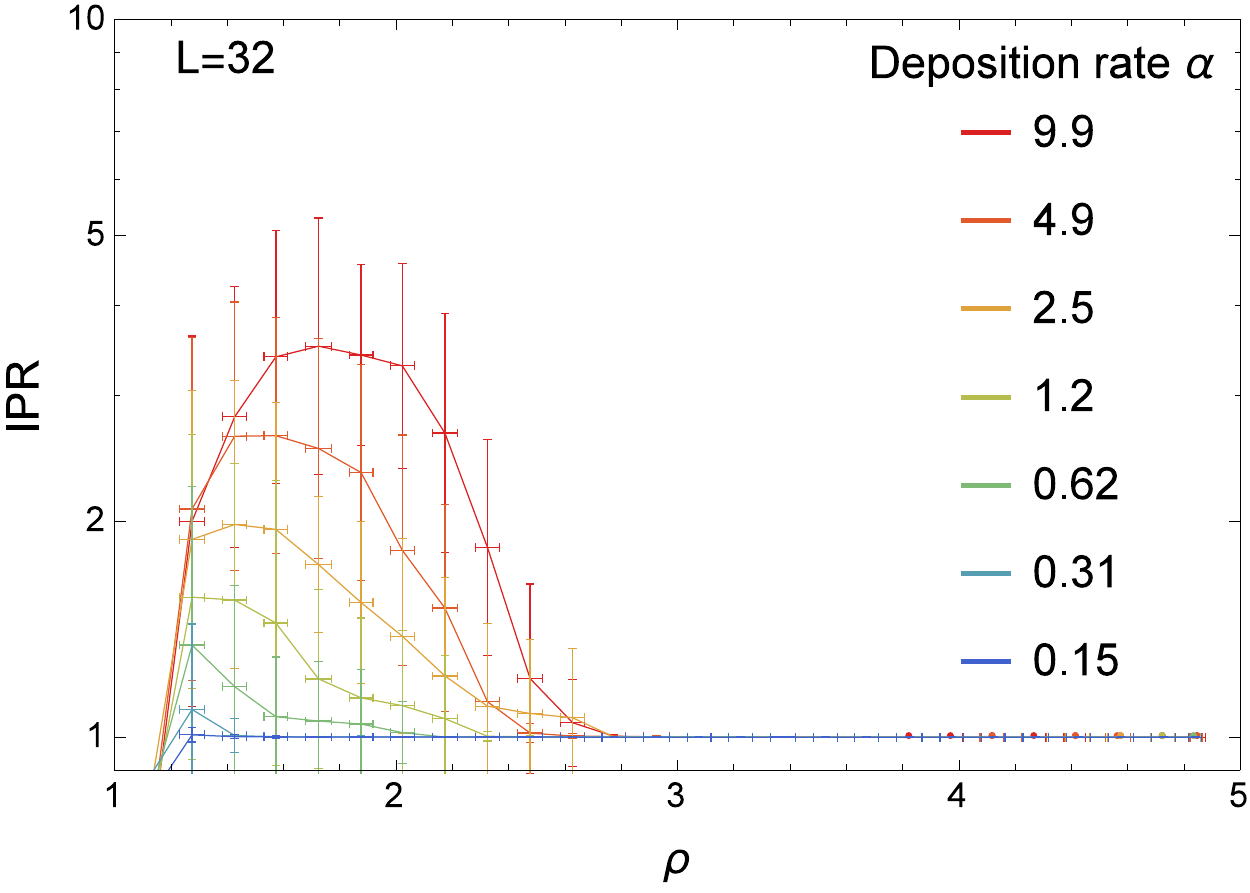}
	\hfill
	\includegraphics[width=0.49\textwidth]{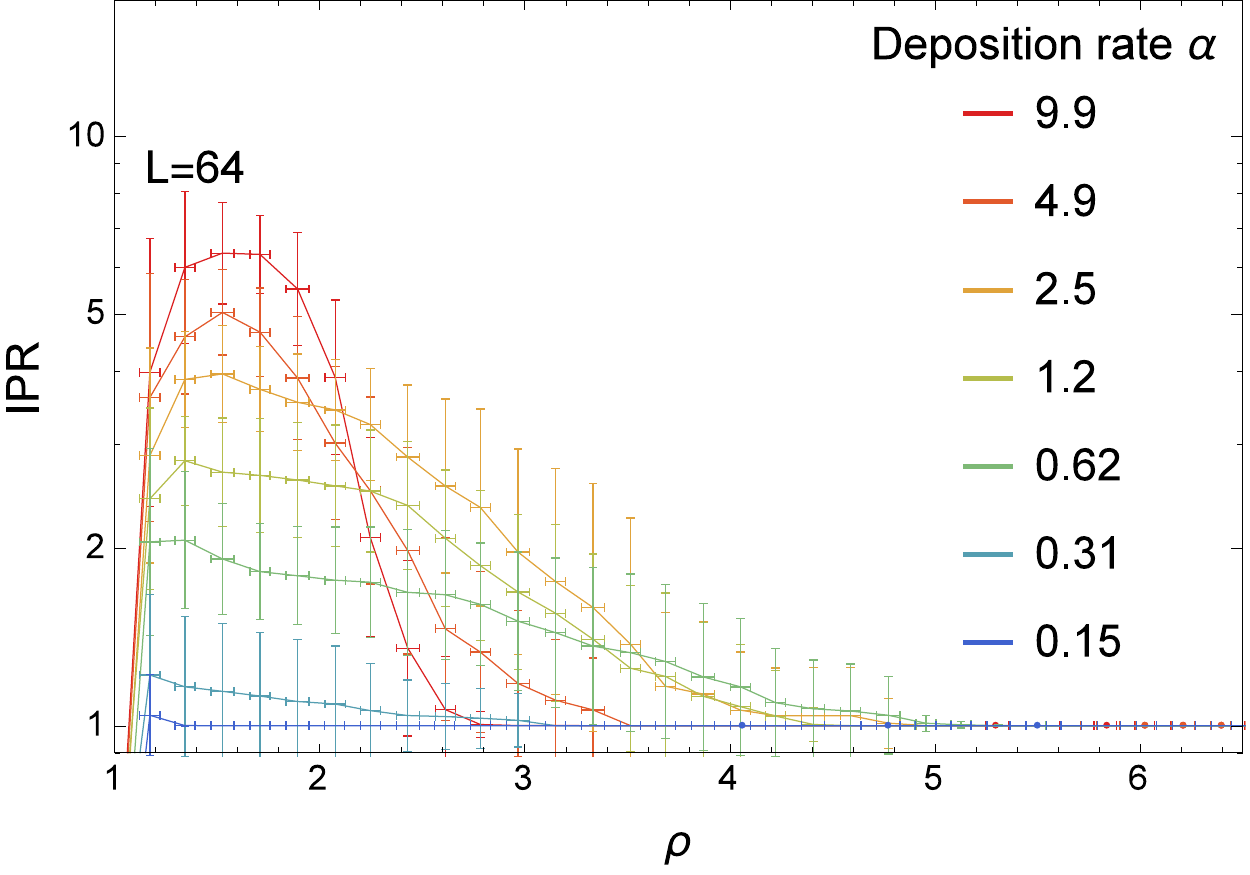}
        \caption{\label{fig:dynamic2Drates} (Colour online) The inverse participation ratio, $\mathrm{IPR}=(\sum_i m_i)^2/\sum_i m_i^2$ (where the sums run over all existing island masses), of condensates formed in $J=1.1, U=3, \sigma=1$ systems with different deposition rates (the fastest is the red topmost curve, the slowest is the blue bottommost one) as a function of the density $\rho$ of the already deposited mass. The mass influxes are given in particles per unit time. The error bars are standard deviations of 20 simulation runs; the $y$-axis is logarithmic. The theoretical estimate $\alpha_{\rm sep}\propto D/L^2$ for the rate separating regimes with one and many condensates yields $0.13$ and $0.033$ for $L=32$ and $64$, respectively. Assuming the proportionality factor 5, the estimated critical densities are $0.65$ and $0.165$, respectively, and they can be seen to separate well the curves for which the IPR remains very close to 1 for all densities (times) and for which it is larger than 1.
		}
\end{figure}

We can estimate the magnitude of the deposition rate $\alpha_{\rm sep}$ that separates the two regimes as follows. We consider only what happens after first $\lfloor\sigma\rfloor$ layers have been filled because this is when condensation begins. A newly added particle stays on the surface of the top-most layer and performs a random walk with diffusion constant $D$ (see Sec.~\ref{sec:dynamics}) until it collides with another particle and becomes the seed of a new cluster. Let us denote (with a slight abuse of notation) the quasi-steady state density of such isolated particles by $\rho$ \footnote{This new $\rho$ should not be confused with $\rho=M/L^2$ defined previously.}. This excludes particles from the complete layers as well as particles in the clusters. If we neglect spatial correlations, the probability that our particle collides with another one during the next step is $\rho$ for $\rho\ll 1$. The probability that the particle has not yet collided after $n$ steps is then $(1-\rho)^n$, and the mean number of steps to collision $\langle n\rangle = \sum_n n\rho(1-\rho)^n = (1-\rho)/\rho \cong 1/\rho$. The time to collision is then $\sim 1/(D\rho)$. During this time the particle departs from its starting point by $\langle r\rangle \sim \sqrt{\langle n\rangle} \sim \rho^{-1/2}$. This distance gives the characteristic length scale for spatial separation of clusters of particles. If it is of the order of the spatial extension of the simulation box $L$, only one cluster --- the condensate --- will form in the system. By equating $\langle r\rangle$ and $L$ we obtain the density $\rho\sim 1/L^2$ at which this happens. To relate this density to the deposition rate $\alpha$ we note that deposition must be balanced by the rate with which particles form clusters; since the clusters are relatively narrow, their contribution to the average density of particles can be neglected. This gives us $\alpha = \rho L^2 D\rho$ where $\rho L^2$ is the number of ``free'' particles in the system. Inserting $\rho\sim 1/L^2$ we obtain $\alpha_{\rm sep}\sim D/L^2$. Hence, if $\alpha_{\rm sep}\gg D/L^2$, multiple condensates are present in the system, otherwise there is only one condensate. Figure~\ref{fig:dynamic2Drates} shows the inverse participation ratio (IPR) of the occupation numbers $\{m_i\}$, which approximately corresponds to the number of condensates, as a function of the density $\rho$ of the already deposited mass (proportional to the physical time), for different deposition rates $\alpha$. The figure indicates that the theoretically predicted $\alpha_{\rm sep}$ correctly estimates the critical deposition rate if $\alpha_{\rm sep} \approx 5 D/L^2$, i.e. the proportionality factor is of the order of 5.

\begin{figure}[htbp]
	\centering
	\includegraphics[width=.49\textwidth]{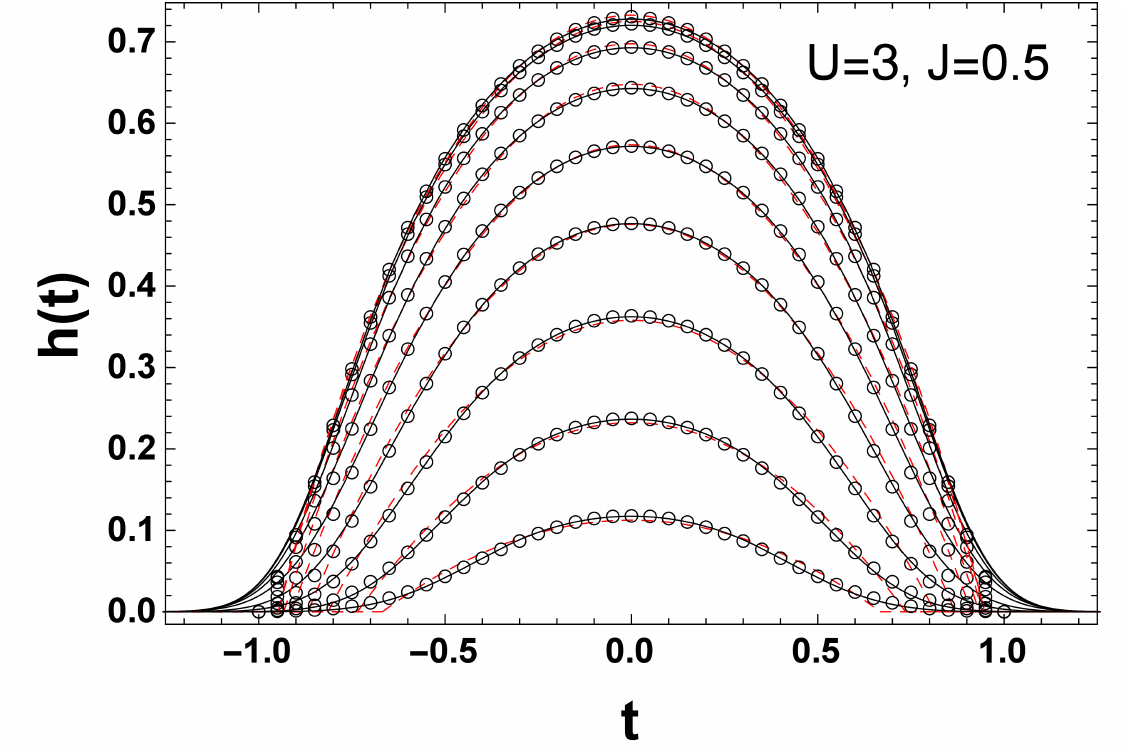}
	\hfill
	\includegraphics[width=.49\textwidth]{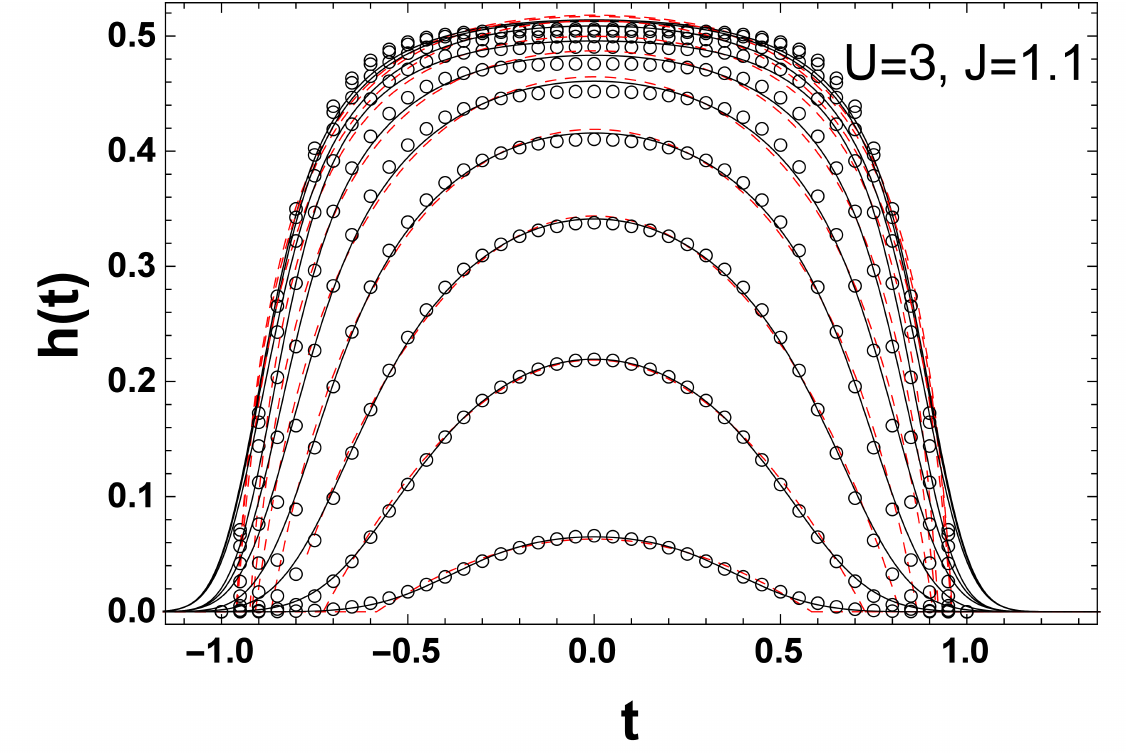}
		\caption{\label{fig:dynamic2Dsec}
		Sections of (2+1)d condensates as in Fig.~\ref{fig:secSide}, but with a constant influx of mass ($0.22$ and $0.62$ particles per unit time for $J=0.5$ and $1.1$, where $\alpha_{\rm sep}\sim D/L^2= 7.6\times 10^{-4}$ and $8.3\times 10^{-3}$, respectively). The circles are sections of condensate snapshots averaged over time and over the multiple condensates, with masses between about 12 -- 34\,000 particles, that formed simultaneously in the system. The red dashed lines correspond to the fitted Eq.~(\ref{eq:shape}), whereas black continuous lines to the analytical shape corrected for the width fluctuations~\cite{BW_JStat} responsible for the tails on the brim of the condensate.
	For details, see~\ref{app:comp}.
		}
\end{figure}

Regardless of whether the deposition rate is high or low, the shape of the condensate(s) can still be well approximated by the equilibrium (1+1)d analytical solution. This is illustrated in Fig.~\ref{fig:dynamic2Dsec}, where we compare the shape obtained in simulations of the (2+1)d LJ model for mass deposition rate $\alpha=0.22$ for $J=0.5$ and $\alpha=0.62$ for $J=1.1$ to the exact solution (\ref{eq:shape}) for the (1+1)d model, with $w,v$ fitted to the cross-sections of the (2+1)d condensate (see~\ref{app:comp} for more details).

\section{Conclusions}

In this work, motivated by thin-film growth processes and, in particular, by the Stranski-Krastanov growth mode, we propose a simple, non-equilibrium physics model in which spatially extended condensates (``islands'') form when the density of particles exceeds a critical value. Our model assumes short-range, valence-bond type interactions between particles, and Lennard-Jones interactions between particles and the substrate on which the growth occurs. Depending on the range $\sigma$ of the Lennard-Jones potential, condensation occurs either directly on the substrate (for $\sigma< 1$) or on a previously formed layer of several particles thick (for $\sigma> 1$).

Although there have been numerous approaches to simulating thin-film growth, (see, e.g., the review~\cite{Jensen}, or a recent kinetic Monte Carlo study~\cite{Smereka}), the most interesting feature of our oversimplified model is that it enables us to calculate many quantities analytically. This is possible due to a pair-factorised steady state (PFSS) probability of microstates in our model. In a (1+1)d version of the model, we have been able to derive the phase diagram of the model, to calculate the critical density for condensation, and to find the shape of the condensate which turned out to depend on the strength of adatom-adatom and adatom-substrate interactions. In the (2+1)d model, which corresponds to the physically relevant growth of 2d layers of adatoms, we have shown that the shape of the condensate is well approximated by the (1+1)d solution. 

We have also studied an open system in which new particles are added at a constant rate. We have shown that condensation occurs above a certain density of particles, and although it is a transient phenomenon, the properties of the condensate are similar to those of the model with mass conservation.

In this work, we have focused on the steady-state, or quasi-steady state properties of the condensate and its late-time dynamics. It would be interesting to broaden our research into the kinetics of initial steps of condensate formation. Further research could also involve manipulating the geometry of the underlying lattice, e.g., introduction of lattice defects which could imitate heteroepitaxial growth more closely~\cite{SKgrowth_strain,SKgrowth_defects}.

\section*{Acknowledgments}
The authors would like to thank Zdzis\l{}aw Burda for valuable discussions. J.O. was supported by the Foundation for Polish Science International Ph.\,D. Projects Programme, co-financed by the European Regional Development Fund covering, under the agreement no. MPD/2009/6, the Jagiellonian University International Ph.D. Studies in Physics of Complex Systems and by the Polish National Science Center grant no. DEC-2013/09/N/ST6/01419 (sec.~\ref{sec:influx} of the paper).
H.N.\ and W.J.\ thanks the german science foundation (DFG) for financial support under Grant No.\ JA~483/27-1 as well as the DFH-UFA graduate school under Grant No.\ CDFA-02-07. B.W. acknowledges the support of the Leverhulme Trust Early Career Fellowship and the Royal Society of Edinburgh Personal Research Fellowship.

\appendix 

\section{Typical values of $\sigma$ for thin-layer growth}
\label{app:atoms}
Assuming that the substrate atoms are uniformly distributed over the lower half-space ($z<0$) of the system, the form of the integrated LJ potential is~\cite{LJ93_RH}
\beq
\epsilon\frac{2\pi}{3}\frac{n\sigma'^3}{d}\left[\frac{2}{15}\left(\frac{\sigma'}{r}\right)^9-\left(\frac{\sigma'}{r}\right)^3\right],
\eeq
where $n$ is the number density of substrate atoms on the surface, $\epsilon$ has the dimension of energy per mol, $\sigma'$ is the range of the LJ potential, and $d$ is the layer spacing of the substrate. The parameters $U$ and $\sigma$ from our formula (\ref{eq:LJmodel}) can be expressed through $n,d,\sigma'$ as:
\beq
	U =  \epsilon \frac{\sqrt{\frac{10}{3}} n \sigma'^3 \pi }{d}, \qquad   \sigma = \left(\frac{2}{15}\right)^{1/6} \sigma'.
\eeq
Carbon or silicon crystals are usually modelled with, very roughly, $3$\AA$<\sigma'<4$\AA~\cite{LJparams1,LJparams2,LJparams3}, whereas lattice constants of C, Si or GaAs are respectively 3.56\AA, 5.43\AA, and 5.65\AA~\cite{Kittel}, which yields $U\approx 0.05$ eV and $\sigma\approx 2.5$\AA. Taking into account that in our work all distances are measured in terms of the lattice spacing $d$, and that $d\approx 1$\AA{} in most metals, $\sigma$ should be about $0.5-3$; these are the values we use in this work. The value of $J$, on the other hand, can be approximated by the Ehrlich-Schwoebel barrier energy, which is typically of the order $0.1-0.5$ eV. Together with $k_{\text{B}} T$ set to $1$ in our simulations and a liquid nitrogen cooled molecular-beam epitaxy temperature of $77$ K, we get the very rough estimates of $U \approx 10, J\approx 50$. In our model, however, we use $J\approx 1$ because for significantly greater values the acceptance rate in MC simulations would become many orders of magnitude smaller, and consequently the simulation times would become unfeasible.

\section{Computer simulations}
\label{app:comp}

To determine the phase diagrams in Figs.~\ref{fig:diagramS1}--\ref{fig:diagramS050930} we used equilibrium Monte Carlo simulations with Metropolis acceptance probability \cite{Metropolis_1953}. A single move consisted of picking up a random site and, if it was occupied, moving a particle to another randomly chosen site anywhere in the system. In comparison to the stochastic simulation of the original dynamics of the model, this significantly reduced the computation time while preserving the stationary state~\cite{BW_JPA2009}. For each pair $(U, J)$, the $64 \times 64$ system (with $\rho=3$ for $\sigma=0.5,0.8,1$ and $\rho=6$ for $\sigma=3$) was simulated for $4\times 10^7$ sweeps\footnote{A ``sweep'' comprises $L$ attempted moves, whereas in (2+1)d it corresponds to $L^2$ attempts.} and, prior to that, it was thermalised for $2\times 10^7$ sweeps.
The strongly rectangular shape of the island for high $U$ and $J$ values is due to the geometry of the square lattice and is independent of the initial conditions.

In Figs.~\ref{fig:2Dshape} and \ref{fig:secSide}, Monte Carlo simulations were performed on a lattice of size $N=200 \times 200$ with $M=25 N= 10^6$ particles and the Lennard-Jones on-site potential. Both cuboid and cylindrical initial condition were used and as we did not find any differences between them, we concluded that thermalisation was long enough to erase any trace of the initial configuration. The simulations took $8\times 10^7$ time steps (around four weeks of computer time), half of which was thermalisation, for the cuboid ($150 \times 150 \times 44$) initial condition, and $4\times 10^7$ time steps, $10\%$ of which was thermalisation, for the cylindrical (diameter $140$, height $60$) initial condition. The final plots shown were obtained from the latter simulation.

The simulations of the dynamics of the (2+1)d model, and the model with mass deposition were performed using a simplified, kinetic Monte Carlo algorithm. Each time step a random site was picked and, if it was non-empty, one of the nearest neighbours was chosen with probabilities $\{r_1,r_2,r_3,r_4\}$ for right, left, top, and bottom jumps, respectively. The particle was then moved between these two sites with probability $u/u_{\rm max}$ where $u$ is the rate from Eq.~(\ref{eq:Um}) and $u_{\rm max}$ was chosen to be larger than the largest possible hop rate for a given set of parameters. This procedure was repeated $L^2$ times. Finally, the physical time was incremented by $dt=1/u_{\rm max}$. In the model with mass deposition, a new particle was added every $dt/\alpha$ steps. This algorithm, although very fast, differs slightly from genuine kinetic Monte Carlo algorithms such as the Gillespie algorithm~\cite{Gillespie_1977}. However, we checked that both algorithms produce indistinguishable results when averaged over a sufficiently long time. Simulations for Figs.~\ref{fig:dynamic2D} and \ref{fig:dynamic2Dsec} were performed on a $128\times 128$ square lattice, with one particle at a randomly chosen site as the initial configuration. For simulations in Fig.~\ref{fig:dynamic2Drates}, with $32\times 32$ and $64\times 64$ square lattice systems, we counted as condensates all clusters both occupying an area greater than $1$ site and having a height greater than $1$ particle. The time to condensation in Fig.~\ref{fig:BW_Tss} was determined as the average time (20-100 simulations per data point) at which the number of clusters larger than $\rho=M/N$ dropped to one for the first time.

The histograms of the (2+1)d condensates in Fig.~\ref{fig:dynamic2Dsec} were obtained from a single simulation run with mass influx $\alpha=0.62$ for $J=1.1$, $\alpha=0.22$ for $J=0.5$. The condensate heights were rescaled according to their masses, the discrete lattice occupations were (linearly) interpolated, and only then were the interpolations averaged producing the histograms.

The simulations of the (1+1)d systems with $\sigma=1$ for Fig.~\ref{fig:1Dshape} were performed on $L=2000$ nodes with $M=60000$ particles. The simulations took $8 \times 10^7$ sweeps, with $7 \times 10^7$ sweeps devoted to thermalisation, and $10^7$ for recording the histogram. The theoretical $\rho_{\text{c}}\approx 1.581$ and the actual subtracted background was $\rho\approx 1.599\pm 0.004$ thick for $U=1.5, J=2$ ($\rho_{\text{c}}=\rho=1\pm 0.001$ for $U=4, J=8$). The theoretical height of the condensate was $h(0)\sqrt{M'}\approx 64.84$, and the actual height measured in simulations was $65.03$ ($h(0)\sqrt{M'}\approx 48.02$, simulations: $47.38$).

\section{Critical density}
\label{app:rhoc}

The transition lines shown for the Lennard-Jones potential in Fig.~\ref{fig:comparison1D} were obtained numerically by diagonalising the matrix $g(m,n)$, as in Eq.~\eqref{eq:eigenproblem}. For faster performance, only a 21-element wide band was retained in the matrix (10 elements below and above the diagonal; the furthermost elements are of the order of $\exp\left(-10 J\right)$), but to avoid numerical errors we used a direct banded matrix solver instead of the iterative (e.g., Lanczos) method. The parameter $U$ was sampled at $0.025$ intervals and the parameter $J$ was determined by the bisection method (the last step of size $\Delta J=0.0195$). The points where the critical density $\rho_{\text{c}}$ from Eq.~(\ref{eq:rhocdef}) increased slower than a logarithm of the matrix size were considered to belong to the condensed phase.
The behaviour was classified as either slower or faster than logarithmic by: first, measuring $\rho_{\text{c}}(L)$ for the weight matrix sizes $250, 500, 1000, 2000$; next, fitting a line in $\ln L$ for the first three points, and another one for the last three points; finally, comparing the two slopes and if the second one was lower, classifying a given $U,J$ pair as belonging to the condensed phase.

In order to determine the critical density of particles above which condensation occurs, we simulated the model with fixed $J,U,\sigma$ while varying the density $\rho\equiv M/L$. Each simulation was thermalised prior to measuring the mass $M'$ of the condensate.

\begin{figure}[htbp]
	\centering
		\includegraphics[width=0.49\textwidth]{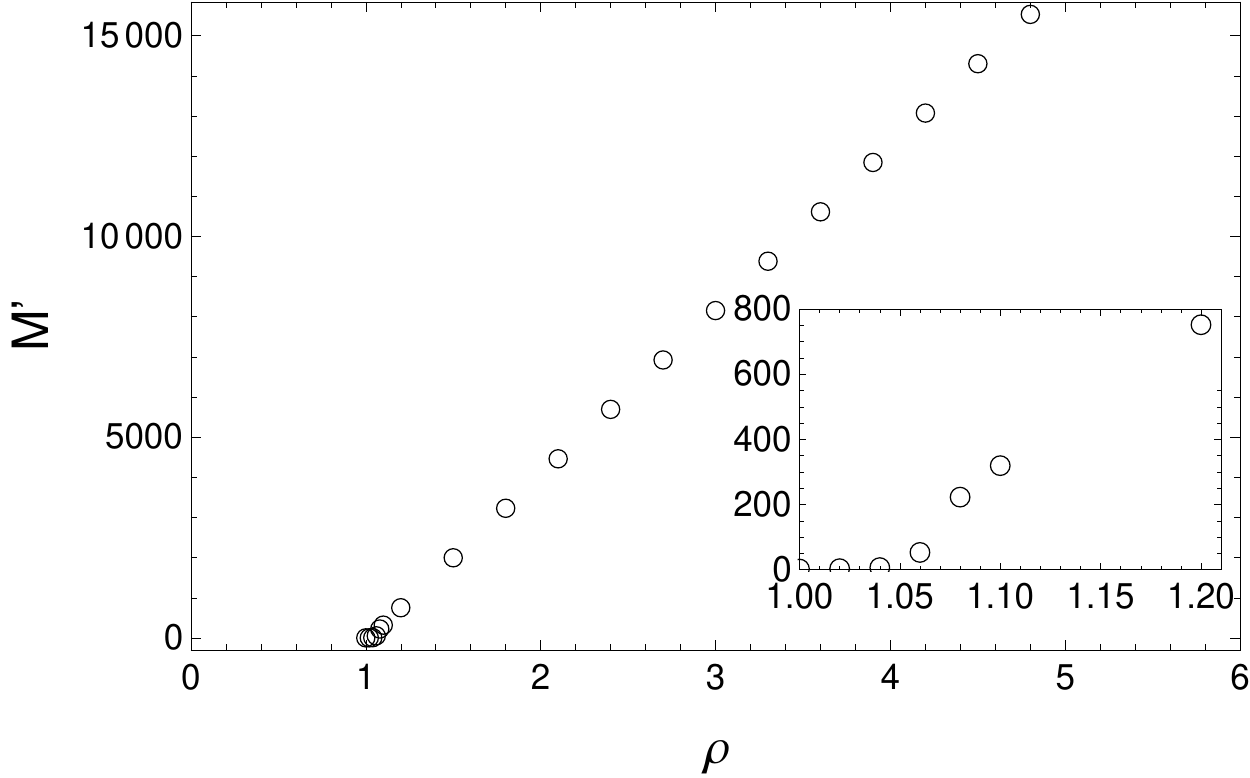}
		\hfill
		\includegraphics[width=0.49\textwidth]{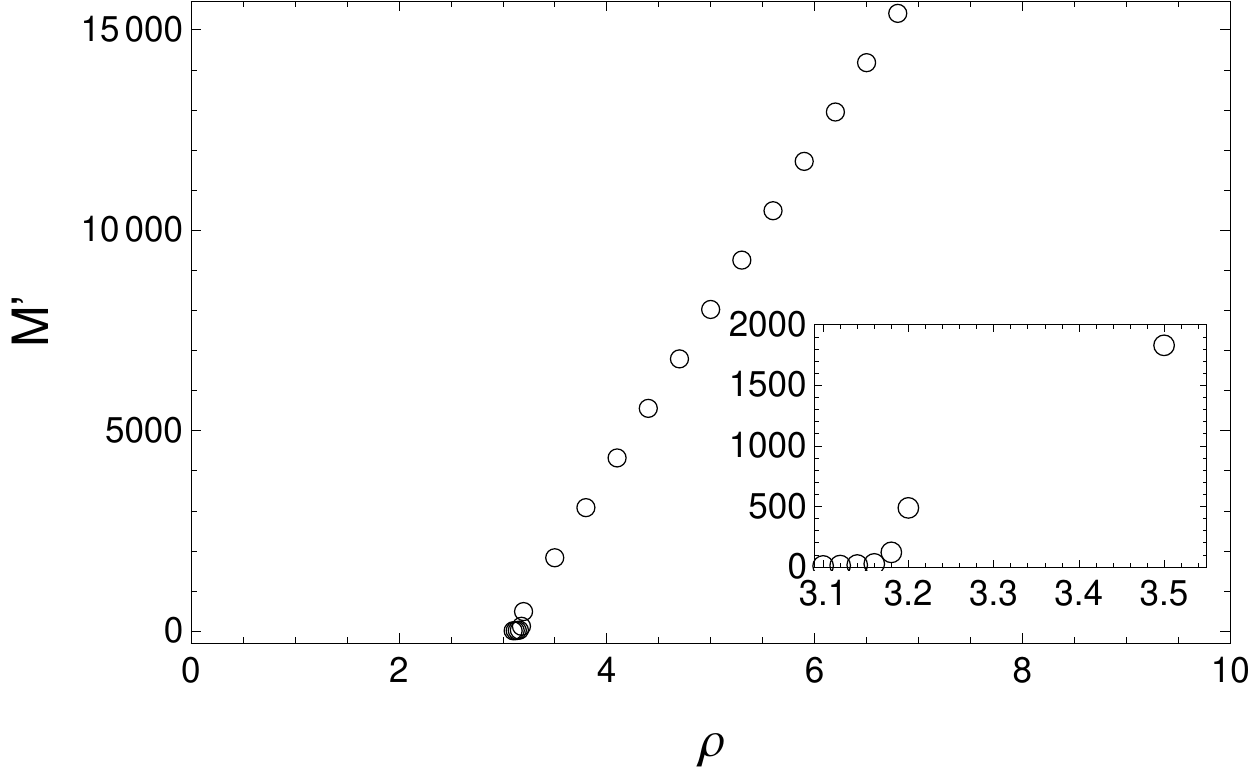}
	\caption{\label{fig:linearDrop} Dependence of condensate mass $M'$ on the mass density $\rho=M/L$ in the (2+1)d systems from Fig.~\ref{fig:densityS1}, $U=3, J=1.1$, with (left) $\sigma=1$ and (right) $\sigma=3$. The insets show details in the small mass and density regime.}
\end{figure}

We then used linear regression $M' = M - L\rho_{\text{c}}$ to determine $\rho_{\text{c}}$ from the sizes of the condensate for different $M$'s, taking into account only sufficiently large $M'$'s, see Fig.~\ref{fig:linearDrop}.

We also performed simulations close to the expected $\rho_{\text{c}}$, as shown in the insets of Fig.~\ref{fig:linearDrop}.
The results indicate that there is a non-linear drop in the condensate mass near $\rho_{\text{c}}$ and hence our method may have produced small but systematic errors when estimating the critical density via linear regression.

\section*{References}

\begin{thebibliography}{10}
\expandafter\ifx\csname url\endcsname\relax
  \def\url#1{{\tt #1}}\fi
\expandafter\ifx\csname urlprefix\endcsname\relax\def\urlprefix{URL }\fi
\providecommand{\eprint}[2][]{\url{#2}}

\bibitem{DDS}
Schmittmann B and Zia R~K 1995 {\em Phase transitions and critical phenomena\/}
  {\bf 17} 3--214

\bibitem{Evans_2000}
Evans M~R 2000 {\em {Brazilian Journal of Physics}\/} {\bf 30} 42 -- 57 

\bibitem{Evans_JPA2005}
Evans M~R and Hanney T 2005 {\em Journal of Physics A: Mathematical and
  General\/} {\bf 38} R195

\bibitem{groskinsky_condensation_2003}
Gro{\ss}kinsky S, Sch{\"u}tz G~M and Spohn H 2003 {\em Journal of statistical
  physics\/} {\bf 113} 389--410

\bibitem{Bialas}
Bialas P, Burda Z and Johnston D 1997 {\em Nuclear Physics B\/} {\bf 493} 505
  -- 516 

\bibitem{ASEP}
Spitzer F 1970 {\em Advances in Mathematics\/} {\bf 5} 246--290

\bibitem{derrida_exact_1993}
Derrida B, Evans M~R, Hakim V and Pasquier V 1993 {\em Journal of Physics A:
  Mathematical and General\/} {\bf 26} 1493 

\bibitem{ASIP}
Reuveni S, Eliazar I and Yechiali U 2011 {\em Physical Review E\/} {\bf 84}
  041101

\bibitem{grosskinsky_condensation_2011}
Grosskinsky S, Redig F and Vafayi K 2011 {\em Journal of Statistical Physics\/}
  {\bf 142} 952--974 

\bibitem{cao_dynamics_2014}
Cao J, Chleboun P and Grosskinsky S 2014 {\em Journal of Statistical Physics\/}
  {\bf 155} 523--543 

\bibitem{Evans_JPA2004}
Evans M~R, Majumdar S~N and Zia R~K~P 2004 {\em Journal of Physics A:
  Mathematical and General\/} {\bf 37} L275

\bibitem{hirschberg_motion_2012}
Hirschberg O, Mukamel D and Sch{\"u}tz G~M 2012 {\em Journal of Statistical
  Mechanics: Theory and Experiment\/} {\bf 2012} P08014 

\bibitem{godreche_condensation_2012}
Godr{\`e}che C and Luck J~M 2012 {\em Journal of Statistical Mechanics: Theory
  and Experiment\/} {\bf 2012} P12013 

\bibitem{godreche_urn_2007}
Godr{\`e}che C 2007 From {Urn} {Models} to {Zero}-{Range} {Processes}:
  {Statics} and {Dynamics} {\em Ageing and the {Glass} {Transition}\/} ({\em
  Lecture {Notes} in {Physics}\/} no 716) ed Henkel M, Pleimling M and
  Sanctuary R (Springer Berlin Heidelberg) pp 261--294 ISBN 978-3-540-69683-4,
  978-3-540-69684-1

\bibitem{daga_phase_2015}
Daga B and Mohanty P~K 2015 {\em Journal of Statistical Mechanics: Theory and
  Experiment\/} {\bf 2015} P04004 

\bibitem{Evans_PRL2006}
Evans M~R, Hanney T and Majumdar S~N 2006 {\em Phys. Rev. Lett.\/} {\bf 97}(1)
  010602 

\bibitem{BW_PRL}
Waclaw B, Sopik J, Janke W and Meyer-Ortmanns H 2009 {\em Phys. Rev. Lett.\/}
  {\bf 103}(8) 080602

\bibitem{SOS1}
Chui S~T and Weeks J~D 1981 {\em Phys. Rev. B\/} {\bf 23}(5) 2438--2441

\bibitem{SOS2}
Burkhardt T~W 1981 {\em Journal of Physics A: Mathematical and General\/} {\bf
  14} L63 

\bibitem{SOS3}
van Leeuwen J and Hilhorst H 1981 {\em Physica A: Statistical Mechanics and its
  Applications\/} {\bf 107} 319 -- 329 

\bibitem{BW_JPA2009}
Waclaw B, Sopik J, Janke W and Meyer-Ortmanns H 2009 {\em Journal of Physics A:
  Mathematical and Theoretical\/} {\bf 42} 315003

\bibitem{BW_JStat}
B~Waclaw J~Sopik W~J and Meyer-Ortmanns H 2009 {\em J. Stat. Mech.\/}  P10021

\bibitem{Meakin_book}
Meakin P 1998 {\em Fractals, scaling and growth far from equilibrium\/} vol~5
  (Cambridge university press)

\bibitem{Krug2}
Krug J 2002 {\em Physica A: Statistical Mechanics and its Applications\/} {\bf
  313} 47 -- 82 

\bibitem{venables_introduction_2000}
Venables J 2000 {\em Introduction to Surface and Thin Film Processes\/}
  (Cambridge University Press) ISBN 9780521785006

\bibitem{Qdots}
N{\"o}tzel R 1996 {\em Semiconductor Science and Technology\/} {\bf 11} 1365

\bibitem{Qdots1}
Daudin B, Widmann F, Feuillet G, Samson Y, Arlery M and Rouviere J 1997 {\em
  Physical Review B\/} {\bf 56} R7069--R7072

\bibitem{kandel_microscopic_1996}
Kandel D and Kaxiras E 1996 {\em Physical Review Letters\/} {\bf 76} 1114--1117

\bibitem{ehrenpreis_numerical_2014}
Ehrenpreis E, Nagel H and Janke W 2014 {\em Journal of Physics A: Mathematical
  and Theoretical\/} {\bf 47} 125001 

\bibitem{LJ93_WS}
Steele W~A 1973 {\em Surface Science\/} {\bf 36} 317--352

\bibitem{LJ93_RH}
Hentschke R 1997 {\em Macromolecular theory and simulations\/} {\bf 6} 287--316

\bibitem{Polya}
P{\'o}lya G 1921 {\em Mathematische Annalen\/} {\bf 84} 149--160

\bibitem{Jensen}
Jensen P 1999 {\em Reviews of Modern physics\/} {\bf 71} 1695

\bibitem{Smereka}
Baskaran A, Devita J and Smereka P 2010 {\em Continuum Mechanics and
  Thermodynamics\/} {\bf 22} 1--26

\bibitem{SKgrowth_strain}
Schmidt O, Kienzle O, Hao Y, Eberl K and Ernst F 1999 {\em Applied physics
  letters\/} {\bf 74} 1272--1274

\bibitem{SKgrowth_defects}
Sakai A and Tatsumi T 1993 {\em Phys. Rev. Lett.\/} {\bf 71}(24) 4007--4010

\bibitem{LJparams1}
Lithoxoos G~P, Samios J and Carissan Y 2008 {\em The Journal of Physical
  Chemistry C\/} {\bf 112} 16725--16728 

\bibitem{LJparams2}
Jia Y, Wang M, Wu L and Gao C 2007 {\em Separation Science and Technology\/}
  {\bf 42} 3681--3695 

\bibitem{LJparams3}
Jorgensen W~L 1981 {\em Journal of the American Chemical Society\/} {\bf 103}
  335--340 

\bibitem{Kittel}
Kittel C and McEuen P 1996 {\em Introduction to solid state physics\/} vol~7
  (Wiley New York)

\bibitem{Metropolis_1953}
{Metropolis} N, {Rosenbluth} A~W, {Rosenbluth} M~N, {Teller} A~H and {Teller} E
  1953 {\em \JCP\/} {\bf 21} 1087--1092

\bibitem{Gillespie_1977}
Gillespie D~T 1977 {\em The journal of physical chemistry\/} {\bf 81}
  2340--2361

\end{thebibliography}
\providecommand{\noopsort}[1]{}\providecommand{\singleletter}[1]{#1}%
\providecommand{\newblock}{}

\end{document}